\documentclass{article}
\usepackage{amsthm, color}
\usepackage{amsmath,amsbsy}
\usepackage{times}
\usepackage{bm}
\usepackage{amssymb, amsfonts}

\usepackage{graphicx,color}
\usepackage{subfigure}

\newcommand{\bmom}{\bm{\omega}}
\newcommand{\bmnab}{\bm{\nabla}}
\newcommand{\bg}{\begin{equation}}
\newcommand{\ee}{\end{equation}}

\newcommand{\ela}[1]{\label{eq:#1}}

\newcommand{\uv}{\boldsymbol{u}}
\newcommand{\xv}{\boldsymbol{x}}

\def\mco{\mathcal{O}}

\newcommand{\Av}{\boldsymbol{A}}

\newcommand{\Rv}{\boldsymbol{R}}

\newcommand{\eps}{\epsilon}

\newcommand{\del}{\nabla}

\newcommand{\er}[1]{$(\ref{eq:#1})$}

\def\u1h{\hat{U_1}}
\def\u2h{\hat{U_2}}

\def\Av{\boldsymbol{A}}

\def\del2{\nabla_2}

\def\tv{\boldsymbol{t}}

\def\<{\langle}
\def\>{\rangle}

\graphicspath{%
    {converted_graphics/}
    {/}
    {C:/Users/Steve/Desktop/Sing2014/}
}

\newcommand{\Iv}{\boldsymbol{I}}

\newcommand{\nb}{\bar{n}}

\newcommand{\GG}{\mathrm{G}}
\newcommand{\HH}{\mathrm{H}}
\newcommand{\eee}{\mathrm{e}}

\usepackage{graphicx}
\begin{document}
\title{A filamentary cascade model of the inertial range}       
\author{Stephen Childress\\Courant Institute of Mathematical Sciences\\New York University\\Andrew D. Gilbert\\Department of Mathematics\\University of Exeter}
       

\date{\today}          
\maketitle
\begin{abstract}
This paper develops a simple model of the inertial range of turbulent flow, based on a cascade of vortical filaments. A binary branching structure is proposed, involving the  splitting of filaments at each step into pairs of  daughter filaments with differing properties, in effect two distinct simultaneous cascades. Neither of these cascades has the Richardson-Kolmogorov exponent of 1/3. This bimodal structure is also different from  bifractal models as vorticity volume is conserved. If cascades are assumed to be initiated continuously and throughout space we obtain a model of the inertial range of stationary turbulence. We impose the constraint associated with Kolmogorov's four-fifths law and then adjust the splitting to achieve good agreement with the observed structure exponents $\zeta_p$. The presence of two elements to the cascade is responsible for the nonlinear  dependence  of $\zeta_p$ upon $p$.

A single cascade provides a model for the initial-value problem of the Navier--Stokes equations in the limit of vanishing viscosity. To simulate this limit we let the cascade continue indefinitely, energy removal occurring in the limit. 
We are thus able to compute the decay of energy in the model.
\end{abstract}

\section{Introduction} \label{intro}
In the limit of vanishing viscosity, and in three space dimensions, the nonlinearity of the Navier--Stokes equations leads to intense vortex stretching and the possibility of a cascade to small scales in both space and time. This cascade is a cornerstone of the structure of stationary homogeneous turbulence at large Reynolds number. It occurs primarily within the inertial range, where the flow is essentially obeying Euler's equations, and the energy imparted at large scales cascades down to a point where viscosity acts to dissipate it.

According to the early work of Richardson on dispersal in the atmosphere and, later, Kolmogorov's 1941 theory, denoted  by K41 (see  \cite{frisch} and references therein), within the inertial range energy cascades down to small scales through a series of steps. At each step the ``eddies'' of one scale break up completely into eddies of a smaller scale. In K41 phenomenology the volume of eddies is conserved in the process. The cascade starts with eddies of size $L$ and  typical velocity $U$. The transitions are taken to be independent of size, leading to the similarity scaling for length $r = \lambda^n L$ at the $n$th step,  where $0<\lambda < 1$. Let  $\delta u(r) $ be  the velocity characteristic of eddies of size $r$. We think of $\delta u$ as the velocity difference within an eddy  seen by an observer moving with the flow which carries the eddy.  The kinetic energy of the $n$th stage eddies is $\sim (\delta u)^2 $ per unit volume. Here $\sim$ indicates a proportionality, since $\delta u $ is a velocity scale of a self-similar vortical structure. We assume that these eddies are created from $n-1$--stage eddies in a time $\sim r / \delta u(r) $. In K41 the flux of kinetic energy $\varepsilon$ is taken as a constant in the inertial range; thus  $\varepsilon \sim (\delta u)^3 /r$ and so $\delta u(r) \sim (\varepsilon r)^{1/3}$.

\begin{figure} 
  \centering
  \includegraphics[bb=0 0 420 315,width=2in,height=1.5in,keepaspectratio]{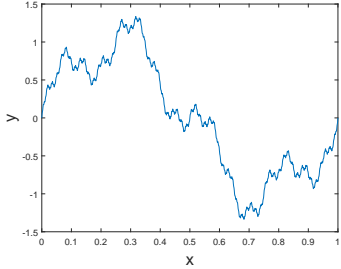}
  \caption{The function $y=\sum_{n=0}^5 (2/3)^n \sin (2\pi 4^n x)$.}
  \label{fig:iteratedsin}
\end{figure}

In the case of stationary fully developed turbulence in three dimensions the above scenario is misleading in one important aspect. The eddies of all sizes are actually superimposed, all extending over a domain of the same volume, the eddies of a certain size  being revealed only by variations on that scale. A one-dimensional representation, which might be thought of as a signal from a hot wire moving though a large eddy, is shown in figure \ref{fig:iteratedsin}.

The Richardson--Kolmogorov  cascade permeates the phenomenology of turbulence theory, despite the fact that the make-up of the structures in mind, the ``eddies'' of the fluid, is unclear. Observation, numerical experiments, and some theoretical constructs suggest that vorticity is concentrated into sheets and tubes in fully developed turbulence. However the complexity of the flow offers little insight into what would constitute an infinite cascade in the sense of Richardson--Kolmogorov. Lundgren has proposed stretched tubular vortices as a source of the cascade and obtained the Kolmogorov $-5/3$ spectrum by analysis of such structures \cite{lund}. The recent observation that elliptical instability of vortex tubes can produce a cascade to smaller scale tubes provides a compelling argument that the mechanism is directly linked to vortex instability \cite{McK}. These experiments also suggest that the cascade resembles the K41 scenario; see \cite{McK2}.

There are however important departures from the Richardson--Kolmogorov  cascade, associated with intermittency of the turbulent dissipation \cite{frisch}. If one considers an average of $(\delta u)^p$, $p\geq 1$ over structures of size $r$ in a stationary field of fully developed turbulence, K41 predicts this should 
scale as $ r^{p/3}$. In fact there are significant departures from experimental results for $p > 3$, as we shall indicate below. Various cascade models have been proposed to correct for this discrepancy. We mention in particular the ``beta'' model, which sacrifices the volume preserving  feature of the cascade, leading to a fractal eddy structure. This and other ``multi-fractal'' generalizations  bring the exponent of $r$ in these averages into better agreement with experiment, but are still without
a physical model of the fluid dynamics of the cascades involved; see \cite{frisch},  chapter 8.

In this note we propose a simple cascade model based upon tubular, essentially filamentary  vortices. This work is an outgrowth of an earlier study which sought to give a vortical interpretation of the beta model \cite{chilvort}.  In brief, the the basic structure of our model is an helical vortex tube, whose core consists of smaller helical vortex tubes on many scales. Our model is unusual in that there will be two Richardson--Kolmogorov cascades involved, neither of which involves the K41 velocity exponent of 1/3. Nevertheless we obtain near-K41 scaling for velocity in an average sense. Since we shall conserve vorticity volume, our cascade is space filling. We shall also find that the model can easily realize the observed intermittency corrections to K41.

In addition to vorticity volume, our model will respect the kinematics of  geometry, and will be constrained to satisfy reasonably well the conservation of kinetic energy. The model is a ``toy'' in that the vortical structures constructed are organized into a clean geometrical hierarchy. In real turbulence such structures would be deformed and unrecognizable. However it is clear from many simulations that filamentary vortices are seen among the small scales of a turbulent flow, and the observations of McKeown {\it et al.}  \cite{McK,McK2} are consistent with a filamentary  hierarchy.
   
One reason for the  well-organised hierarchy of our model is that it bypasses the dynamics of vortical interactions. The only  dynamical constraint imposed is that associated with Kolmogorov's four-fifths law \cite{frisch}. Consider the volume average $\<(\delta u)^p\>$ for any integer $p\geq 1$. The exponents $\zeta_p$ are defined by $\<(\delta u(r)/U)^p\>\sim (r/L)^{\zeta_p}$. The four-fifths law implies that $\zeta_3=1$, this being one of the few exact results for the inertial range. For discussion and a rigorous proof as a local property of weak solutions of the Euler equations see \cite{eye1,DR}.  We emphasize that here the constraint is imposed on a single realization of the cascade, and in that sense it is being taken as a local property of cascading eddies according to Euler's equations. However we must pay a price for the restricted dynamical input. Our cascade will comprise a two parameter family. One of these parameters can be fixed by bringing the $\zeta_p,\; 1\leq p \leq 10$ into agreement with experimental data. The agreement is remarkably good given that only one parameter is varied. Other constraints will be discussed below to further narrow the choice of cascade.

We shall not be displaying experimental results for the $\zeta_p$. We shall refer instead to a simple formula for $\zeta_p$ obtained by  She and Leveque \cite{SL}, which agrees very well with experimental results for $\zeta_p$ in the range $1\leq p\leq 10$:
\bg
 \zeta_p = \frac{p}{9}+2\Big[1-\Big({2\over 3}\Big)^{p/3}\Big].
\label{eqsl}
\ee
This formula will be the ``target'' relation for our model: agreement with the She--Leveque values is regarded as reasonable agreement with observation, at least for $p\leq 10$.
This formula was derived for stationary turbulence on the basis  of certain hypotheses concerning the ratios of moments of the dissipation. It is noteworthy that the assumptions are based on the idea that the dominant dissipation occurs in filamentary vortex structures.

\section{Formulation of a binary branching model} \label{form}

We are interested in filamentary vortex structures with finite kinetic energy.  One might hope, when peering into the inertial range at extremely large Reynolds number, to find that Navier--Stokes turbulence might exhibit some local order. That is, given a certain time window, an observer moving with the local mean velocity might be able to see a few well structured steps of the Richardson--Kolmogorov cascade.
Globally, the complexity is immense, remnants of the cascade having been distorted by long range vortical interactions. Our attempt here is to model this local order. We need a structure with well-defined length scales allowing the possibility of a self-similar ordering, as well as a way of defining an ``eddy''. We also need a way in which each structure can break into two (or more) substructures, in a manner that does the least damage to the geometry and topology of vortex lines. 

The family of structures we use to do this, to build an idealised model, uses helices. The equation of one turn of a straight helix wound on the $x$-axis is 
$(x,y,z)= (b\cos t,b\sin t, ct) , 0 \leq t\leq 2\pi$. Here we call $b$ the \emph{turn radius}, and $2\pi c$ is the \emph{pitch}, equal to the advance along the axis of one turn of the helix. The length of one turn is $2\pi \sqrt{b^2+c^2}$, and its \emph{slope} is $c/b$. Thus in our construction these helices can have varying radii, pitch and can be wound around any curve, including another helix. Also, as in the duplication of DNA, a helix can easily split into two sub-helices without having any intersections. In our model, any one vortex filament  of our cascade will be taken as a tube wound into a helical structure, the tube having a circular core of constant vorticity and carrying a certain circulation $\Gamma$. 
To secure finite energy, the vortex tube can be regarded as  wound helically around a closed helical curve whose shape will be determined by the cascade. We will thus be  dealing with   an ``iterated helical tube''. An eddy associated with the filament will be one turn of this helical structure. 

\begin{figure} 
  \centering
 (a) \includegraphics[scale=0.28]{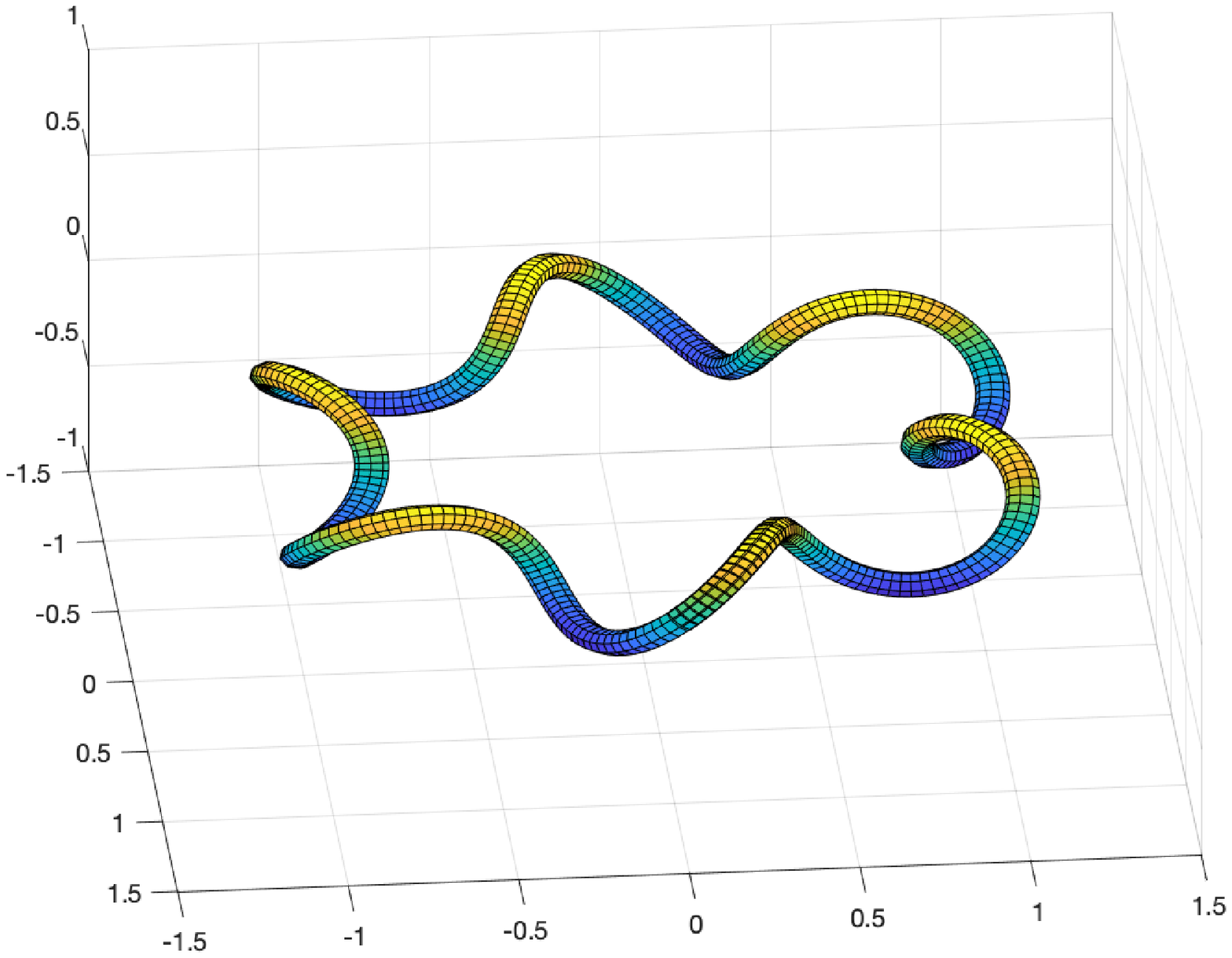}
 (b) \includegraphics[scale=0.28]{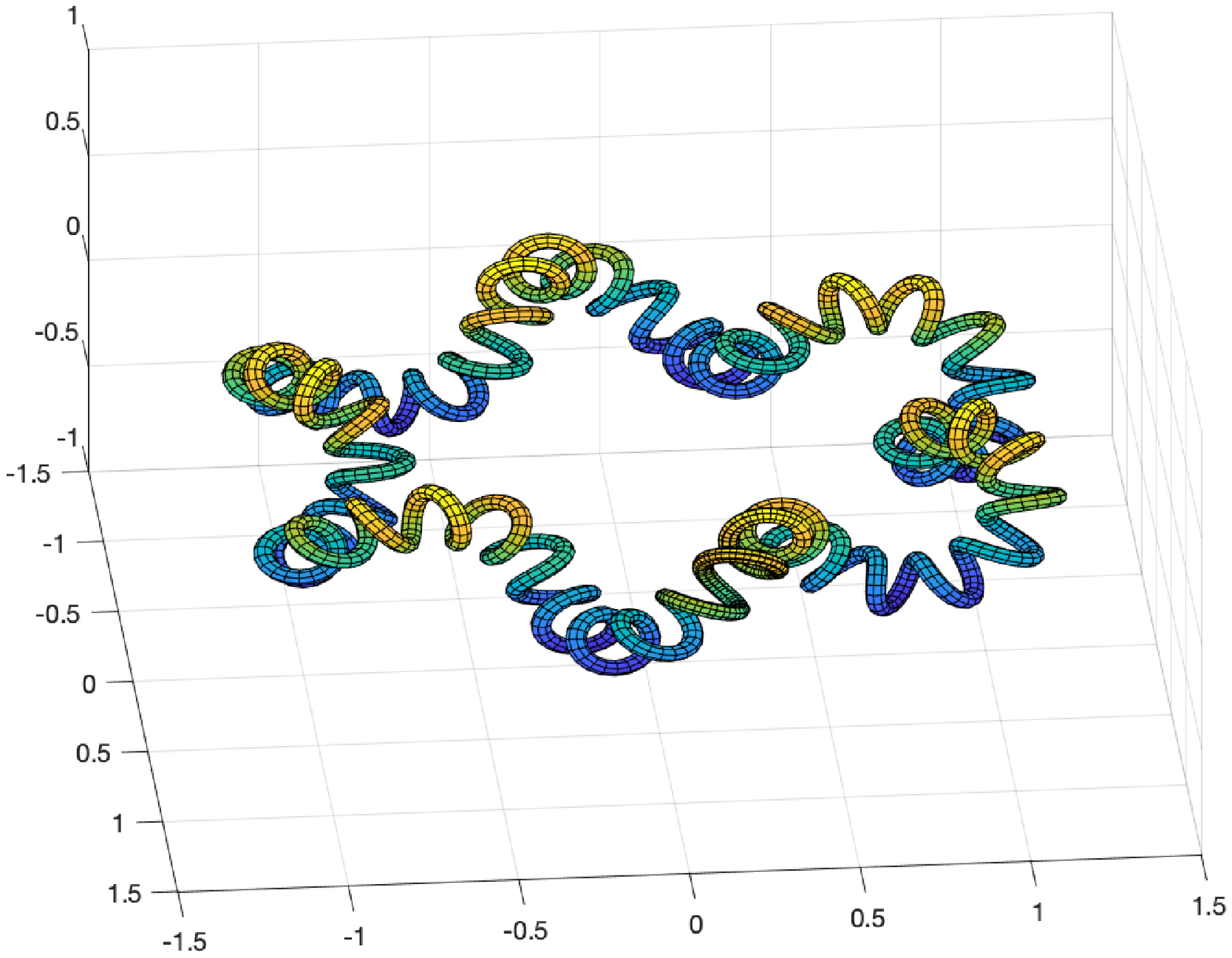}
  \caption{(a) A helix with $m=6$ turns wrapped around a circle and parameters defined as in \er{helixgeom} with turn radius $b=0.2$, pitch $2\pi c$, $c=1$, core radius $r=0.05$, (b) a helix with 60 turns wrapped around the helix in (a), with turn radius $b=0.1$, $c\simeq 0.1$, core radius $r=0.03$.
  } 
  \label{fig:simplehelix}
\end{figure}

As examples, figure \ref{fig:simplehelix} shows two building blocks. In (a) a helix with 6 turns is wound around its centreline, a circle. This helix could, through instability, throw off one or more helices wrapped around it, as in (b) which shows a helix with 60 turns whose centreline is the helix in (a). Clearly such a process involves a new smaller length scale, and stretching of vorticity. At the same time averaging over this smaller scale reveals the presence of the larger scale as well.
This is a property of a once iterated helical filament.  If we require that a single turn (eddy) of the helix in (b) is geometrically similar to a single turn (eddy) in (a), in other words in terms of length, core radius, turn radius and pitch, we have geometrical constraints on our  modelling. These constraints will  introduce self-similar structures into the cascade.

\begin{figure} 
  \centering
 (a) \includegraphics[scale=0.25]{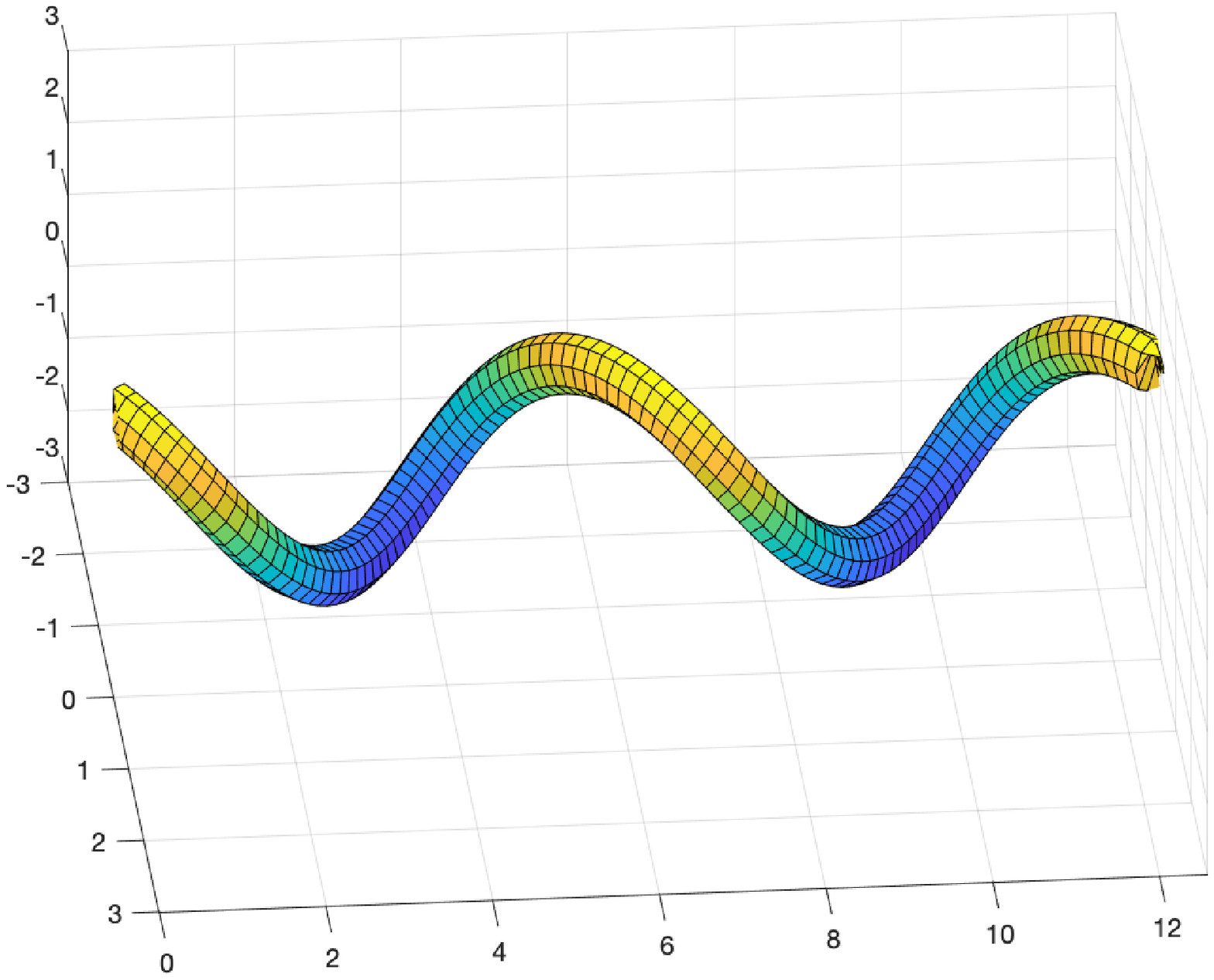}
 (b) \includegraphics[scale=0.25]{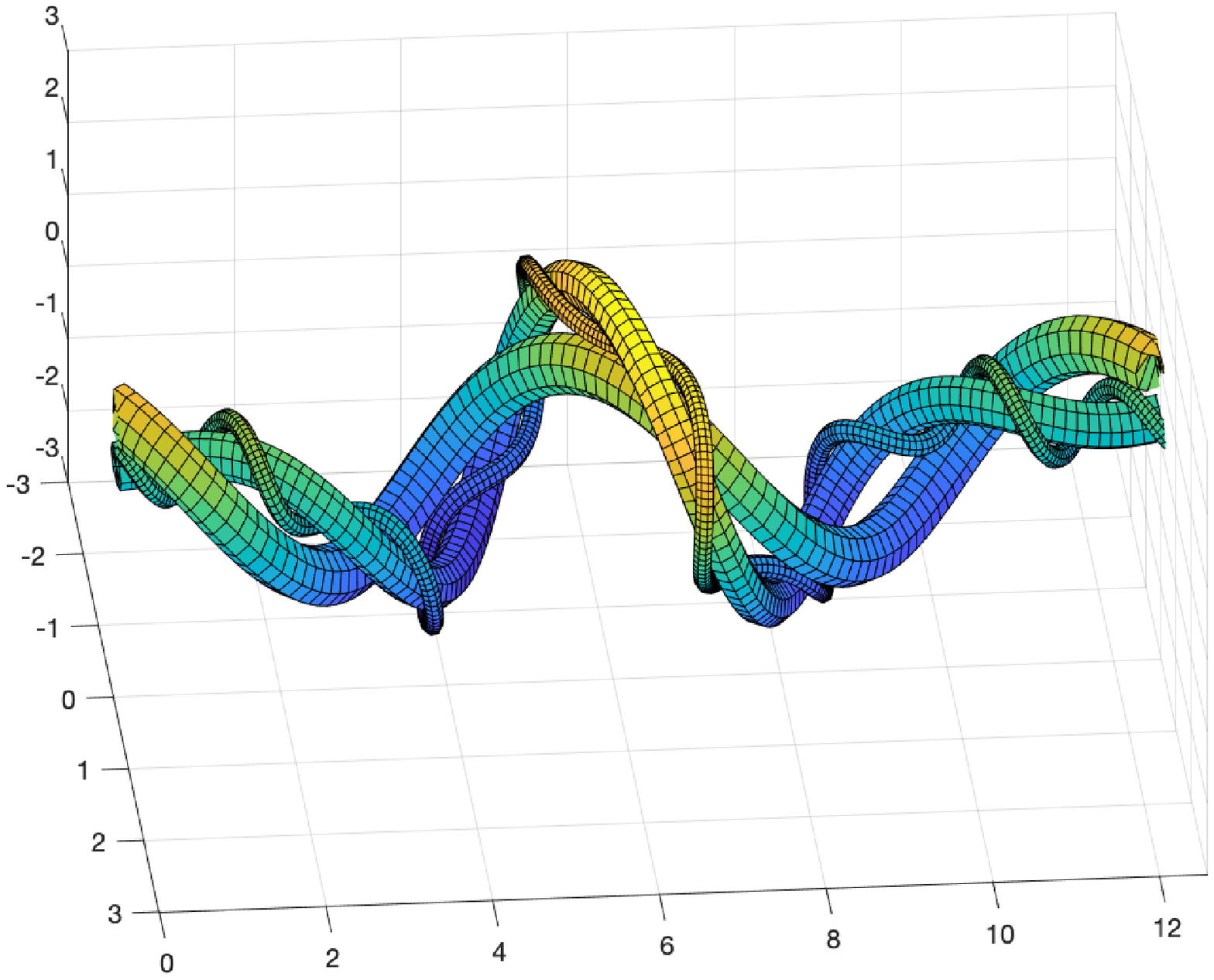}\\
 (c) \includegraphics[scale=0.25]{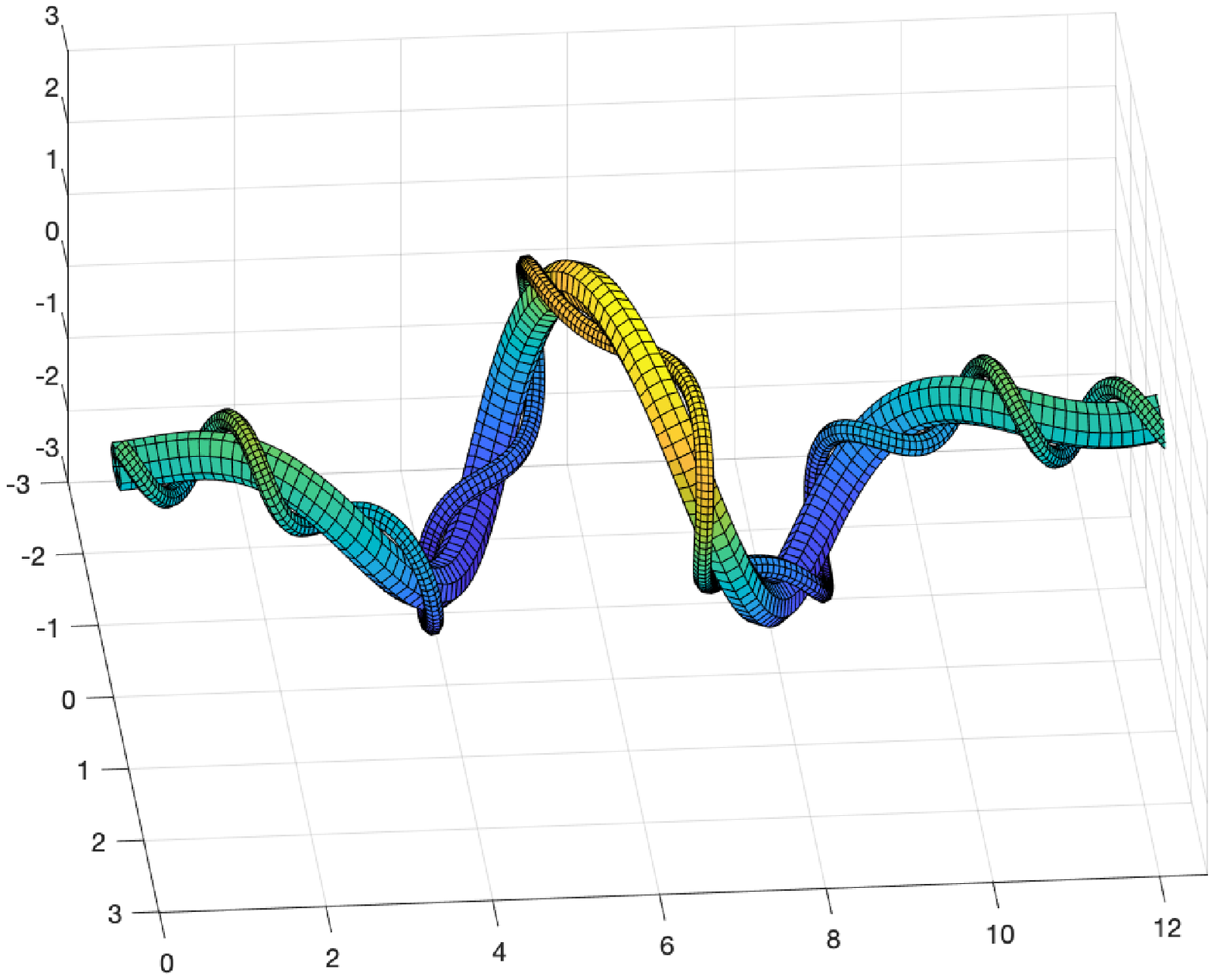}
 (d) \includegraphics[scale=0.25]{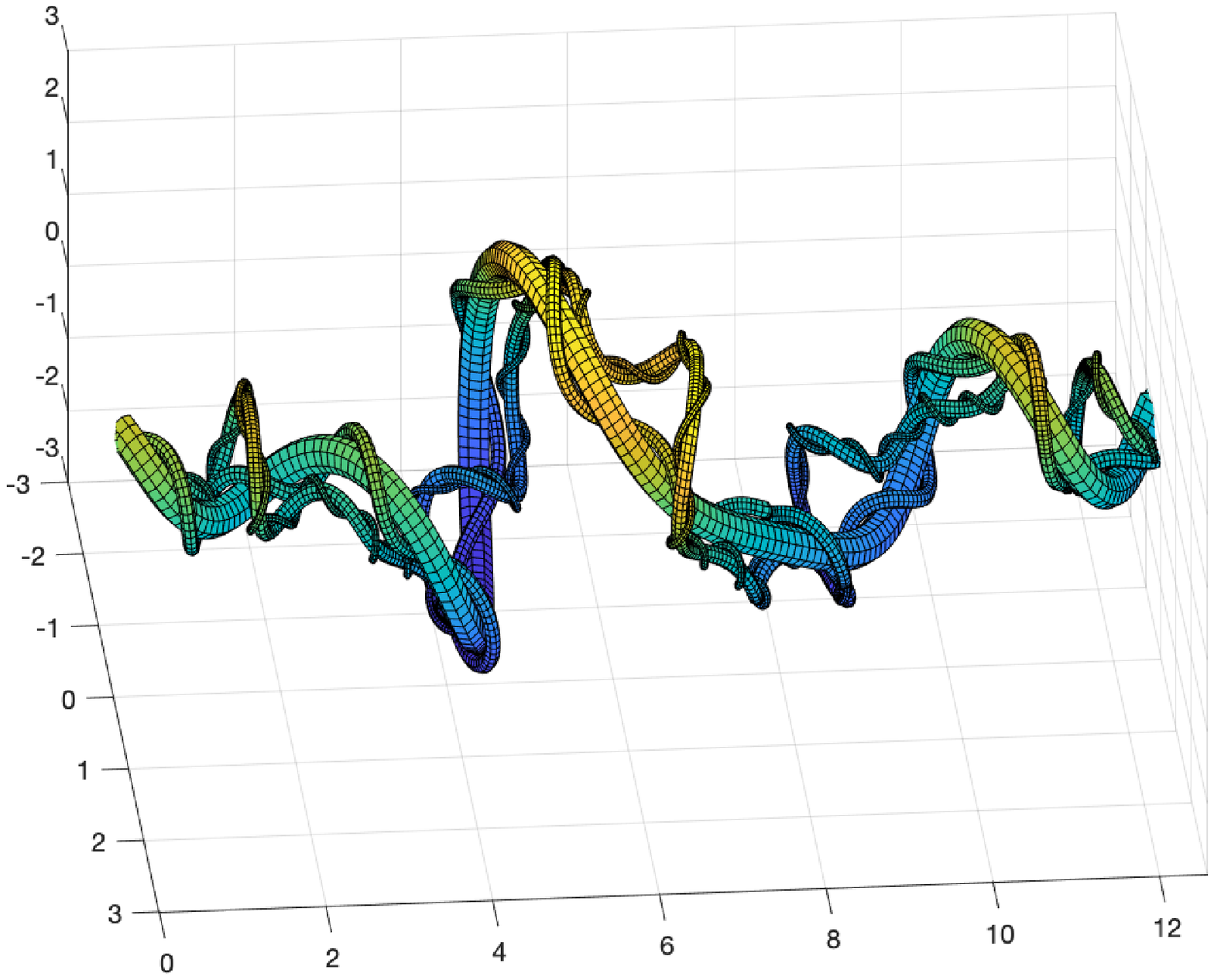}\\
 (e) \includegraphics[scale=0.25]{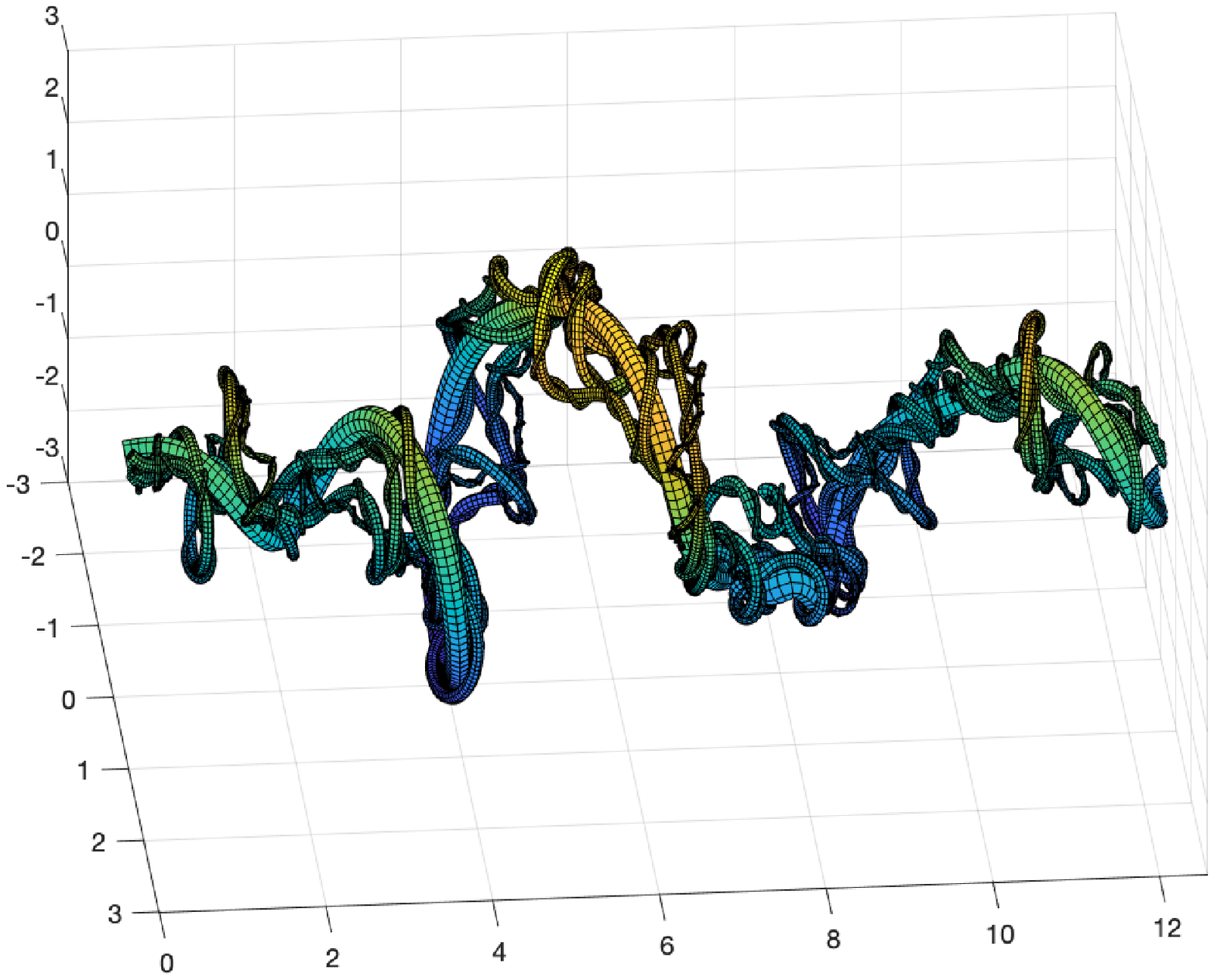}
  \caption{(a) A piece of root helix, which in (b,c) splits into two daughter filaments [0], [1], with different amounts of stretching and different scaling factors. To show the winding of the filaments, in (b) both the parent ``ghost'' filament and the two daughter filaments are shown, while in (c) the parent is removed to leave the daughters. In (d) further splittings create four granddaughter filaments, and in (e) eight great granddaughters. Note that the splittings will not generally occur at the same times when the filaments in (d) and (e) are generated.}
  \label{fig:cascadehelix}
\end{figure}

To produce a model of a cascade to ever finer scales, we now develop this basic idea and consider in detail the {\em splitting} of a  helix into two distinct helices, each carrying a fixed  circulation, the sum of the two being the circulation $\Gamma$ of the unsplit filament.  It will be useful to think of our filaments as comprised of vortical strands. The splitting divides the filament into two distinct sets of strands. Each will be similar in geometry to the starting helix, but not necessarily scaled in the same way, an important distinction. This splitting is assumed to be instantaneous and the two daughter filaments to twist about one another but follow the path of the unsplit helix, a process that involves stretching and intensification of vorticity. It is the twist in the daughter helices which will give rise to a new iteration of the  helix. This process is illustrated in figure \ref{fig:cascadehelix} in which a piece of helix in (a) splits to form the two daughter helices in (b): these are scaled differently and so have different widths and different numbers of turns, but both have centrelines that are the original helix in (a). After some time each of the the daughter filaments will each split instantaneously into two granddaughter filaments that twist about the daughter helices, the two splittings generally occurring at different times. These are shown in figure  \ref{fig:cascadehelix}(c), and a further step in (d).  This process is then continued indefinitely, producing, after $n$ steps, $2^n$ filaments of smaller and smaller circulation, and smaller and smaller length scales. However, the structure will preserve all scales of the cascade. Each splitting event will occur at a  definite time which we shall specify in due course (the figures show the levels in the hierarchy, not a snapshot in time). We shall also need to specify how kinetic energy is distributed and flows down the cascade of scales.

\subsection{The geometry of splitting}

We now summarize how we will set up and evolve the vorticity within such a program. We consider a single cascade initiated at time zero by a helical filament of circulation  $\Gamma$. To have a finite kinetic energy we take the helical filament  to be closed and wound on a torus as in figure \ref{fig:simplehelix}(a). This is a bit of an artifice since we are interested in local interactions. We shall deal with the question of finite energy in section \ref{model}.
 
The equation for the points of the initial filament centreline in Cartesian coordinates is given by
\bg
 \Rv(t)=( (R+b\cos t)\cos(t/m), (R + b \cos t)\sin(t/m), b\sin t),\quad 0 \leq t\leq 2\pi m.
\ela{helixgeom}
\ee
This helix makes $m$ turns on the torus and $R=mc$ is the radius of the torus. We use $b$ and $c$ as the parameters of the helix and its length is approximately $2\pi m \sqrt{b^2+c^2}$, even though it is wrapped around a curve.
The circular core of the filament, containing the vorticity, is taken to have core radius $r$, and a parameter $\eps=\tfrac{1}{2} r/\sqrt{b^2+c^2}$ will occur in our analysis of the cascade. We shall denote this initial helical structure by H and refer to it as the \emph{root filament} of our cascade. 

We  describe the the splitting of filaments using a binary notation. H will split into two helices  H[0] and H[1]; we call these the [0] helix  and the [1] helix. The circulations of these two filaments will be  $\Gamma[0]=\beta_0\Gamma$ and $\Gamma[1]=\beta_1\Gamma$ with $\beta_0 + \beta_1 = 1$. Note that $\beta_0$ is the volume fraction of the the [0] helix relative to the root, since at the moment of splitting the two daughters have the length of the parent.  Following the splitting, we assume that each daughter filament is stretched uniformly, by factors $s_0$ and $s_1$ respectively. These stretched filaments will also be closed filaments, with the [0] filament wrapped around the [1] filament. 
We emphasize that all vorticity is accounted for in the splitting. Our cascade preserves vorticity volume and hence does not produce a fractal (but does give what is sometimes referred to as a \emph{fat fractal}). Three levels of splitting are depicted in figure \ref{fig:cascadehelix}.

It is at this stage that binary self-similarity is introduced. We specify that one turn of H[0] will be a copy of one turn of the root helix, but smaller by a factor $\lambda_0$ (in all dimensions, including the core size), and similarly for H[1] with factor $\lambda_1$. Since in general these scaling factors will not be the same, our model begins to differ significantly from the beta model and the general phenomenology of simple (non-statistical) K41 scaling.

Indeed, taking the stretching to be uniform, it is immediate from the scaling of the core radii that 
\bg
 \lambda_0=\sqrt{\beta_0/s_0}, \quad \lambda_1= \sqrt{\beta_1/s_1}. 
\ee
After stretching and establishment of the similitude, our structure  will have $m[0] = {ms_0/ \lambda_0}$ turns of the [0] helix and  $m[1]={ms_1/ \lambda_1}$ turns of the [1] helix.
(Although in general $m[0]$, $m[1]$ will  not be integers, they will be large compared to one and the nearest integer will be taken as they play a minor role in the computations below.)
The [0] helix now has length $2\pi m \sqrt{b^2+c^2}s_0$, since it has resulted from stretching the root helix by the factor $s_0$, and has parameters $b[0]$, $c[0]$. The [1] helix will have parameters $b[1]$, $c[1]$. As  we have noted, the [0] helix may be regarded as  wrapped around the the [1] helix. 

We approximate this ``helix'' around a helix'' as a ``helix around a large circle of equal length.'' Then we have, approximately
\bg
m[0]= {2\pi m \sqrt{b^2+c^2}s_1\over 2\pi c[0]}={2\pi m \sqrt{b^2+c^2}s_1\over 2\pi\lambda_0 c}= {s_0\over \lambda_0}m.
\ee
Therefore $s_0  > s_1$ and
\bg
 {b\over c}= \sqrt{\left(\frac{s_0}{s_1}\right)^2-1}.
\ela{boverc}
\ee
Thus $b/c$, the inverse slope of the helices, is fixed  from the stretching parameters throughout our structure, $2\pi b > 2\pi c$ expressing the excess length needed to turn a line into a helix.

We shall also specify that that the pair of daughter helices, H[0], H[1] are together oriented along the  path occupied by  what we shall describe as the ``ghost'' of the root helix H --- the root helix is no longer there, but the the daughter helices spiral about the region it previously occupied. 
This is important for the retention of the scales of variation of the vorticity and velocity fields as the cascade proceeds. In effect the pair of daughter helices should be regarded as merged into a restoration of the parent, when viewed on the scale of the parent. One way to achieve this is to take the [1] helix as wound on the ghost of the root helix. Then we would 
have $2\pi m[1]c[1]=2\pi m \sqrt{b^2+c^2}$ or $s_1=\sqrt{1+(b/c)^2}$. We have above also taken the [0] helix as wound around the [1] helix: the figure \ref{fig:cascadehelix}(b) depicts this configuration with the ghost of the parent from (a) shown, while in (c) only the two daughters are shown.   Thus, combined with
\er{boverc} we then obtain the geometrical constraint
\bg
s_1^2=s_0.
\ela{geomconstr}
\ee
If, similarly $s_0^2=s_1$, then the H[1] may wrap around H[0]. We shall later see how well our cascade satisfies these constraints and use them to fix parameters of the model.

We are using the term ``helix'' rather loosely here, to describe a structure each of whose components consists in the small of helical turns. The only time we need to be specific about the geometry is in the calculation of the local transfer of energy, as discussed in the next section, as well as in the imposition of \er{geomconstr}.
We emphasize that in visualizing this cascade it must be kept in mind that 
our helices are filamentary, i.e. slender vortex tubes. The degree of slenderness is set by the root helix.

To assess the effect of  binary splitting, consider how the velocity scales with eddy size. Since vorticity increases in each filament by the stretching factor, we have characteristic velocities $U[0]=\sqrt{s_0\beta_0}\;U$, $U[1]=\sqrt{s_1 \beta_1}\;U$, where $U$ is the characteristic root velocity. Writing
\begin{equation}
U[0]/U=\sqrt{s_0\beta_0}=\lambda_0^{\alpha_0}, \quad
U[1]/U=\sqrt{s_1\beta_1}= \lambda_1^{\alpha_1}, 
\label{eqalphasdef} 
\end{equation}
we evaluate for $s_0=1.6, \; s_1=1.22,\;  \beta_0=0.25$, numbers which will appear below, and obtain $\alpha_0=0.49$, $\alpha_1=0.145$. These values must be compared to the  K41 scaling exponent of $1/3$. No single cascade step  of the present model, for example from root to H[0] or root to H[1], will yield this scaling. But we shall see that the present model does give  $\zeta_1=0.348$ for the above parameter values, close to K41.  This emphasizes the importance of averaging over all elements of a cascade which combines two scaling factors. The point we highlight is that the resulting single scaling exponent is a statistical quantity, a fact which is widely recognized but not often emphasized when discussing phenomenology.

\begin{figure} 
  \centering
  \includegraphics[bb=0 0 396 339,width=4in,height=3.42in,keepaspectratio]{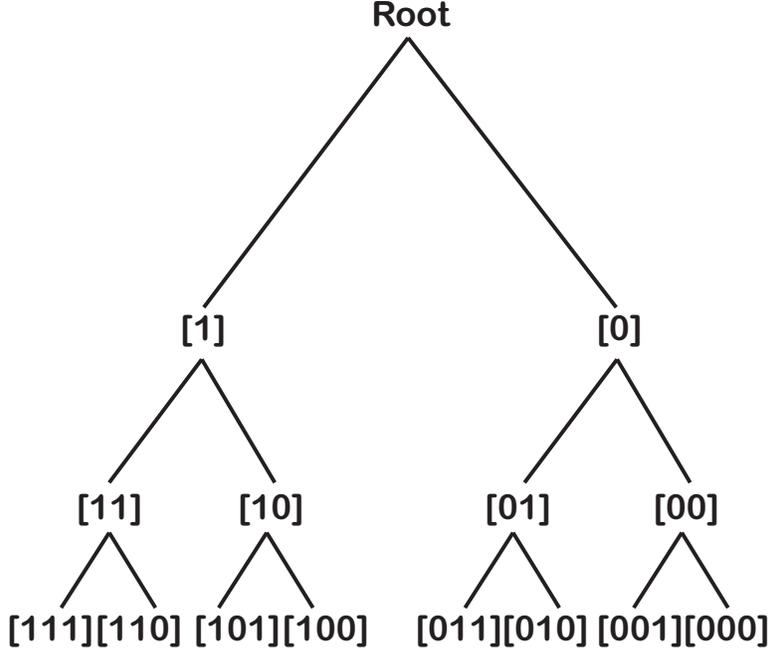}
  \caption{The binary cascade.}
  \label{fig:Bigtree}
\end{figure}

The cascade now continues by splitting of the two daughters H[0], H[1] into four granddaughter helices. The pair H[01], H[00] splits off from H[0], and H[10] and H[11] from H[1]. Here and below the digits read from left to right give the sequence of splittings,  H[10] being wrapped around H[11].  The scale factors $\lambda_0,s_0$ apply to derive H[10] from H[1], the scale factors $\lambda_1,s_1$ similarly give H[11] from H[1]. We show in figure \ref{fig:Bigtree} these and subsequent steps in the binary cascade, mirroring the structures in physical space in figure \ref{fig:cascadehelix}.

\section{Kinetic energy of filaments} \label{energy}

We are interested in the energy transfers and energy conservation during the idealised model cascade that we have outlined above. In order to facilitate these calculations, In this section we  focus on the energy of a single helical filament, and of a pair that are intertwined. The kinetic energy $E$ of a compact vortical structure in $\scriptstyle{\mathbb{R}}^3$ may be computed from
\bg
 E={1\over 8\pi}\iint {\bm{\omega}(\xv)\cdot\bm{\omega}(\xv^\prime)\over |\xv-\xv^\prime|}\,dV\,dV^\prime.
\ela{enercalc}
\ee

\subsection{A single helix}

We shall apply this formula first to the closed, filamentary,  helical structures of our cascade. Consider first the filament with axis given by \er{helixgeom} and core radius $r$.
The calculation of energy uses a classical regularization to deal with the singularity of the energy of vortex line. We give details in  appendix A. The energy consists of two parts, external and internal. Since the filaments are slender, the external part is independent of the distribution of vorticity in the core. It can therefore be computed assuming that vorticity is concentrated at the core boundary.  The result is
\begin{align}
  E_{\mathrm{ext}} \approx {\Gamma^2 m\over 2}\bigg[ \int_\eps^\infty \bigg(  &   {b^2\cos\psi+c^2\over ({4b^2\sin^2 \tfrac{1}{2} \psi+ c^2\psi^2})^{1/2} }  -{{b^2}\cos\psi +c^2 \over c\psi}\bigg) d\psi
 \notag\\
 &  \qquad\qquad  -{b^2\over c}\, {\rm Ci}(\eps)+{c\over 2}\int_{\eps / m}^\pi { \cos\psi\over  \sin \tfrac{1}{2} \psi}\, d\psi\bigg],\quad m\gg 1. 
\end{align}
Here
\bg
\tfrac{1}{2} \int_{\eps / m}^\pi { \cos\psi\over  \sin \tfrac{1}{2} \psi}\, d\psi=-\log \tan(\eps/4m)-2\cos(\eps/2m), 
\ee
and 
\begin{equation}
\mathrm{Ci}(z) =  - \int_z^\infty \frac{\cos t}{t} \, dt 
\end{equation}
is the cosine integral. 

%
%
For constant core vorticity the internal energy is
\bg
E_{\rm int} = {m\Gamma^2\sqrt{b^2+c^2}\over 8}
\ee
to leading order.

\subsection{The interaction energy between two helices}

We now consider the interaction energy between the two helices  H[0] and H[1] that result from a splitting. This is obtained from \er{enercalc} when the $\xv$ integration is over one filament and the $\xv^\prime$ integration is over the other. Since the [0] helix is wound on a [1] helix, and the dimensions of the two, determined by the scaling factors, are of comparable order, we have a complicated intertwining of curves. We shall model it by replacing the H[1] by a ring of the same length, which then serves as the axis on which H[0] is wound. 
We denote the helix with subscript `$\mathrm{h}$' and the ring with subscript `$\mathrm{r}$'. Thus we seek to compute
\bg
E_{\rm interact}= {\Gamma_{\mathrm{h}}\Gamma_{\mathrm{r}}\over 4\pi}\int_0^{2\pi m}
\!\!\!
   \int_0^{2\pi m} {\tv_{\mathrm{h}}(t)\cdot\tv_{\mathrm{r}}(t^\prime)\over |\Rv_{\mathrm{h}}(t)-\Rv_{\mathrm{r}}(t^\prime)|}\,dt\,dt^\prime,
\ee
where
\begin{align}
 & \Rv_{\mathrm{r}}(t)=(mc\cos (t/m),mc\sin (t/m) , 0), \\
 &  \Rv_{\mathrm{h}}(t)=((mc+b\cos t)\cos(t/m), (mc + b \cos t)\sin(t/m), b\sin t),\\
 & \tv_{\mathrm{h}}(t)\cdot\tv_{\mathrm{r}}(t^\prime)=-bc\sin t \sin [(t-t^\prime)/m]+ (c^2+(bc/m) \cos t) \cos[(t-t^\prime)/m], \ela{tangs}\\
 & |\Rv_{\mathrm{h}}(t)-\Rv_{\mathrm{r}}(t^\prime)|=\big[b^2+(2mbc\cos t+2m^2c^2)\big(1-\cos[(t-t^\prime)/m]\big)\big]^{1/2}, \ela{Rs}
\end{align}
for $ 0 \leq t\leq 2\pi m$. We will drop the term involving ${bc/ m}$ in \er{tangs} and $2mbc$ in \er{Rs} as negligible at large $m$.

At this point the calculation proceeds similarly to the direct energy calculation, and is in fact simpler. We can divide into inner and outer contributions as follows:
\begin{align}
   E_{\rm interact}\approx{ \Gamma_{\mathrm{h}}\Gamma_{\mathrm{r}}\over 2\pi}\int_0^{2\pi m}dt^\prime\bigg[ & \int_0^{Am^\alpha}{c^2\over\sqrt{b^2+c^2(t-t^\prime)^2}}\, d(t-t^\prime)
\notag\\
  + & \int_{Am^{\alpha-1}}^\pi {-b\sin (t^\prime + m\psi)+c\cos\psi\over 2\sin \tfrac{1}{2} \psi}\, d\psi\bigg].
\end{align}
We see that the $t^\prime$ term does not contribute under $t^\prime$ integration. Thus we obtain
\bg
E_{\rm interact}\approx  \Gamma_{\mathrm{h}}\Gamma_{\mathrm{r}} m c\Big[\log{2cAm^\alpha\over b}-\log{Am^{\alpha-1}\over 4}-2\Big]=\Gamma_{\mathrm{h}}\Gamma_{\mathrm{r}} m c\Big[\log{8m c\over b}-2\Big].
\ee

We are now able to compute the energy (external plus internal plus interaction)  of our system of H[0] and H[1]. Let the root energy, involving $m$ turns of the root helix, be  $E$. Let H[0], involving $m[0]$  turns, have energy $E[0]$ in isolation, and similarly $E[1]$ for H[1]. Finally let $E[0,1]$ be the interaction energy of $m[0]$ turns of H[0] in the presence of a ring filament with the properties of H[1], and  whose length is that of H[1]. Then the ratio of total energy of this system to the root energy is given by

\bg
E_{\rm total}/E = s_0\beta_0^2 E[0]/E+s_1 \beta_1^2 E[1]/E+s_0\beta_0 \beta_1  E[0,1] /E.
\ela{totener}
\ee        
Here the factors of $\beta_{0,1}$ are the scaling factors for circulation, obtained at the time of splitting, and before any stretching has occurred. The factors $s_{0,1}$ come from a product of the scaling by $\lambda_{0,1}$ in the size of one turn of the helix, and the scaling by $s_{0,1}/\lambda_{0,1}$
of the number of turns of the helix.

We shall make use of this formula in the section \ref{model} to validate the approximate conservation of energy in the splitting of a helical filament.  We note here that every term in the energy balance has the factor $\Gamma m c$ from the root filament. The remaining dependence upon $m$, and only dependence surviving in 
\er{totener}, occurs in terms which contribute a logarithmic divergence. These logarithmic terms also contain the effect of filament slenderness through the parameter $\eps$. As a result the energy per unit turn of a filament and its daughters depends rather weakly on $m,\eps$.

\section{A model of the inertial range in stationary turbulence}\label{model}

In this section we develop in detail the averaging of $(\delta u)^p$ over our branching, helical  model of the inertial range. Our aim is calculate the structure functions $\zeta_p$ and to determine the parameters of the model which yield $\zeta_p$ in  good agreement with experiment. We shall also consider the energy balance at a given step of the cascade. To avoid clutter we will replace the normalising dimensional quantities $L$ and $U$ by unity in much of what follows.

In the model the turns of helical filaments  (our eddies) have  scalings of size of the form $\lambda_0^k\lambda_1^{n-k}$ at step $n$, with $0\leq k\leq n$. The individual filaments have well defined lifetimes, of the form $s_0^{-k}s_1^{-(n-k)}$, that is, proportional to the inverse vorticity in each core. The volume of a filament scales as $\beta_0^k\beta_1^{n-k}$. In stationary turbulence, we must be careful to differentiate the lifetime of a filament from the time of existence of a particular scale of variation over the core of the filament. Even as  the splitting has proceeded down to the Kolmogorov scale, the root structure, now involving many internal scales, remains, as do the sub-filaments on all scales. For example in figure \ref{fig:cascadehelix}(d) the region occupied by the great granddaughter filaments still outlines the original root filament in (a). 
In essence we must keep separate the size of eddies and the size of the various structures they comprise. These structures can be identified by sampling of the ensemble at a particular resolution. Once eddies are removed at the dissipation scale, root energy is maintained to sustain the stationary state.  It is at this point that the external supply is manifest, as a renewal of energy in a new root filament.

We assume that all cascades are identical so we may restrict attention to the average over a single cascade. The time of appearance of a filament is not needed to compute our average, but will be of interest below when we consider a model of freely decaying turbulence. Since we are interested in the average of 
$(\delta u)^p$, we note that at step $n$ of the cascade, in each of the various structures labelled by $k$, $(\delta u)^p$ scales  as 
\begin{equation}
(\delta u)^p \sim U^p(\beta_0 s_0)^{pk/2}(\beta_1s_1)^{p(n-k)/2},\; k=0,1,\dots,n . 
\end{equation}

Our goal now is to calculate the structure function 
\begin{equation}
Q_p(r) = \<(\delta u)^p\> \sim r^{\zeta_p}
\end{equation}
as a function of scale $r$, and so the exponent $\zeta_p$, by averaging over our hierarchy of structures with different scales and strengths. Let us select a scale $r$ and consider the contribution to $Q_p(r)$ from structures in our hierarchy: those whose scale $\lambda_0^k\lambda_1^{n-k}$ is approximately $r$ will contribute to $Q_p(r)$. Obviously $r$ is a continuous variable whereas we have a set of discrete scales labelled by $n$, the level in the hierarchy, and $k$ giving the various branches with a suitable binomial weight. 

\begin{figure} 
  \centering
  \includegraphics[bb=0 0 452 280,width=3.5in,height=2.17in,keepaspectratio]{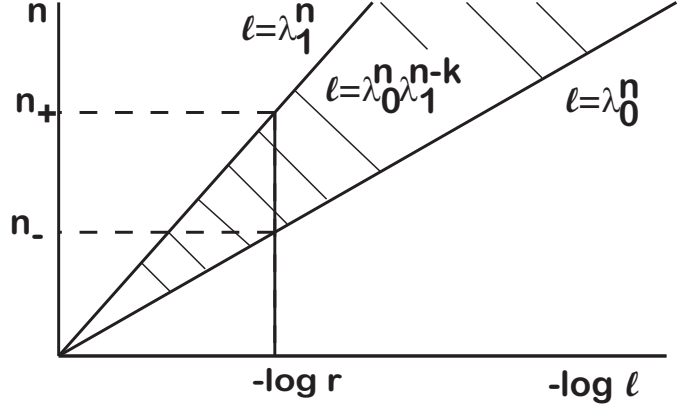}
  \caption{Eddies contributing to an average at a scale $r$ for $\lambda_0  < \lambda_1$. For level $n$ in the hierarchy between $n_- \leq n \leq n_+$ there are contributions from eddies of scale, say $\ell = \lambda_0^k\lambda_1^{n-k}$ for  $k= k(r,n)$ values given by (\ref{eqkeq}).}
  \label{fig:knplotadgillus}
\end{figure}

Let us suppose for definiteness that $\lambda_0 < \lambda_1$ and take a small scale $r\ll 1$. Then for small $n$ all the structure scales $\lambda_0^k\lambda_1^{n-k}$ for varying $k$ will be significantly larger than $r$, and for large $n$ they will all be much smaller. There will be a range of levels $n$, say $n_- \leq n \leq n_+$ for which $\lambda_0^n \leq  r \leq \lambda_1^n$ and at these levels structures for some $k$ values will contribute to the average. Specifically these are the $k$ for which $r \simeq \lambda_0^k\lambda_1^{n-k}$ or equivalently 
\begin{equation}
 k  \simeq \frac{\log r - n \log \lambda_1 }{\log ( \lambda_0/\lambda_1) }\, . 
 \label{eqkeq}
\end{equation}
Typically it would be a range of nearby $k$ values contributing, but a range that would not change as we vary $r$ and $n$, since the scales of the individual contributions to the correlation function go down geometrically. 
Up to the multiplicative constants that we are ignoring in this type of scaling argument, we can take for each $r$ and $n$ the single structure with $k$ rounded in (\ref{eqkeq}) to the nearest integer. The contributions to a given scale $r$ are shown schematically in figure \ref{fig:knplotadgillus}. We need to sum these contributions, over the appropriate values of $k$ and $n$. First though we consider a special situation in which $\lambda_0 = \lambda_1$ and the calculation is easier.

\subsection{The self-similar case $\lambda_0=\lambda_1\equiv \lambda$}

This special case is of interest because of its simplicity. The geometry of the structures is the same whether they are generated by splitting into a [0] daughter or a [1] daughter; however the two branches are still not equivalent as we allow $\beta_0 \neq \beta_1$ and so $s_0 \neq s_1$: structures can be stretched by different amounts and so at each level $n$ of the hierarchy we have structures that have the same scale but different vorticity intensities and so different contributions to the structure function $Q_p(r)$. 
In the following we will often refer to scaling factors by the name of the quantity they scale, again effectively setting the root scales $U$, $L$ equal to unity. The structures of size $r=\lambda^n $ may now be identified with a given step $n$.  These structures are space filling (have the total volume of the root filament), but appear at different times and have different lifetimes, of order $s_0^{-k}s_1^{-(n-k)}$, before breaking into smaller structures. 

We shall compute $\zeta_p$ by imposing the four-fifths condition $\zeta_3=1$. The latter follows from the vanishing of a dissipation scaling parameter at $p=0$, 
which is a consequence of the assumption that the support of dissipation is finite in the limit of zero viscosity; see \cite{SL}. This condition also results from the independence of volume on the scale of the structures involved.

We now have the link $\lambda=\sqrt{\beta_0/s_0}=\sqrt{\beta_1/s_1}$ and we recall that velocity scales like $\lambda s_0$  or $\lambda s_1$ as we create [0] or [1] helices. We compute $(\delta u)^p$ by integrating $(\beta_0s_0)^{pk/2}(\beta_1s_1)^{p(n-k)/2}$  over the volume occupied by the structures of size $r= \lambda^n$, namely $\beta_0^k\beta_1^{n-k}$, multiplying by the number of such structures, and summing over $k$.  Thus we have
\begin{align}
Q_p(r) = \<(\delta u)^p\> & = \sum_{k=0}^n {{n}\choose{k}} \beta_0^k\beta_1^{n-k} (\beta_0s_0)^{pk/2 }(\beta_1s_1)^{p(n-k)/2} \\
& =\big[\lambda^{-p}(\beta_0^{p+1}+\beta_1^{p+1} ) \big]^n\sim  r^{\zeta_p} = \lambda^{n\zeta_p}. 
\end{align}
The four-fifths law then implies 
\bg
\lambda=\lambda^{-3}(\beta_0^{4}+\beta_1^{4}),
\ela{45same}
\ee
or $\lambda=(\beta_0^{4}+\beta_1^{4})^{1/4}$, with $s_{0,1}=\beta_{0,1}/\lambda^2$.
Thus

\bg
\zeta_p={\log( \beta_0^{p+1}+\beta_1^{p+1})\over \log (\beta_0^4+\beta_1^4)^{1/4}} - p .
\ela{eqlts2}
\ee

\begin{figure}[tbp] 
  \   
  \includegraphics[bb=0 0 420 315,width=2.5in,height=1.87in,keepaspectratio]{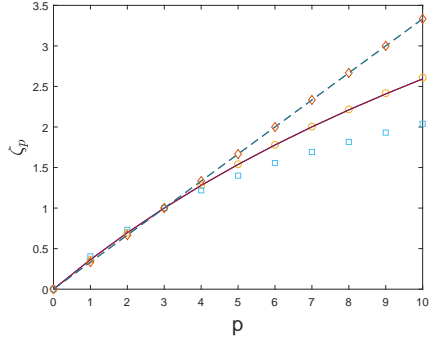}
  \caption{$\zeta_p$ using \er{eqlts2} with optimal $\beta_0 = 0.438$ or $0.562$ (circles), with $\lambda=0.608$. Also we show results for $\beta= 0.4$ or $0.6$ (squares), and $\beta=0 .5$ (diamonds). The solid line is the She--Leveque result, the dashed line K41.}
  \label{fig:lambdasthesame2}
\end{figure}

We show in figure \ref{fig:lambdasthesame2} this result for the optimal $\beta_0=0.438$ or 0.562. Taking $\beta_0=0.562$ we have $s_0=1.52, s_1=1.18$. The other values of $\beta$ shown in the figure illustrate what we shall find occurs more generally. For $\beta$ less than optimal, the $\zeta_p$ are too low. For larger values, the results lie close to K41. What is remarkable is how close we fall to the She--Leveque result at the optimal $\beta$. We note that once the constraint of equal $\lambda = \lambda_0 = \lambda_1$ is relaxed, there will be a second free parameter (which we shall take to be $s_0$). Optimization is then over two parameters.
An important condition of the cascade in stationary turbulence is that the flux of energy be constant. Here, we must sum for each daughter the volume fraction times velocity squared divided by a time, the latter being the inverse vorticity. Thus in terms of scaling factors the quantity
\bg
\bar{\varepsilon} = \beta_0^2s_0^2+\beta_1^2s_1^2
\ee
should be unity. Since here $s_{0,1}=\beta_{0,1}\lambda^{-2}$ we indeed find $(\beta_0^4+\beta_1^4)\lambda^{-4}=1$. Note this constraint is equivalent here to the four-fifth's law \er{45same} since $\lambda^2=\beta_{0,1}/s_{0,1}$

We also point out the connection made here to a single cascade. The case $\beta_0=0.5$ shown in figure \ref{fig:lambdasthesame2} yields K41. But then, again since $s_{0,1}=\beta_{0,1}\lambda^2$, we have $s_0=s_1$. Thus all the  daughter filaments are equivalent.   This is the  only example within our model of a single cascade. It yields K41 scaling for all $p$ with $\beta=1/2, s=\sqrt{2}, \lambda=8^{-1/4}=0.5946$
The [0] helix cannot now wrap around the [1] helix. So a different structure is needed, for example taking both helices wrapped around the centerline of the parent but out of phase by half a turn.

\subsection{The general case, $\lambda_1\neq \lambda_0$}

We now consider the general case where $\lambda_1 \neq \lambda_0$ and structures at each level $n$ of the cascade have varying scales. Bearing in mind the binomial weight and  arguing as above, each set of branches labelled by $k$ and $n$ contributes an amount that scales as   
\bg
 {{n}\choose{k}} \beta_0^k\beta_1^{n-k} (\beta_0s_0)^{pk{/2}}(\beta_1s_1)^{p(n-k){/2}} 
 \label{eqnk1} 
\ee
to $Q_p(r)$ at the scale $r = \lambda_0^k \lambda_1^{n-k}$. We take the cascade to continue to an arbitrary number of levels $n\to\infty$, for this calculation relevant to the inviscid limit. Of course in reality the smallest filamentary scale is fixed at the Kolmogorov scale. We first give an approximate argument that gives numerical values for the scaling exponents $\zeta_p$ for $Q_p(r) \sim r^{\zeta_p}$. We will then refine the discussion below in section \ref{sseclargedeviation}. 

 First, let us generalise to some general measurement $Q(r)$ and let $Q_\lambda$ be the scaling factor linked to an eddy of size scaling factor $\lambda$, so that the contribution in (\ref{eqnk1})  above becomes
\bg
 {{n}\choose{k}} (\beta_0 Q_{\lambda_0})^k (\beta_1Q_{\lambda_1})^{n-k}    .
\ee
If we look at the contribution to $Q$ from \emph{all} structures at level $n$ for all $k$, that is the contribution to $Q(r)$ integrated over all scales $r$, this is 
\begin{align}
(\beta_0 Q_{\lambda_0}+\beta_1Q_{\lambda_1})^n & =\sum_{k=0}^n {{n}\choose{k}}(\beta_0 Q_{\lambda_0})^k (\beta_1Q_{\lambda_1})^{n-k} 
\label{eqQsumnk} 
\\
& = (\beta_0 Q_{\lambda_0}+\beta_1Q_{\lambda_1})^n \sum_{k=0}^n {{n}\choose{k}}f_0^k f_1^{n-k},
\end{align}
where 
\bg
f_0=\beta_0 Q_{\lambda_0}/(\beta_0 Q_{\lambda_0}+\beta_1Q_{\lambda_1}),\quad f_1=\beta_1Q_{\lambda_1}/(\beta_0 Q_{\lambda_0}+\beta_1Q_{\lambda_1}).
\ela{fg}
\ee

Now the inertial range analysis in terms of the exponents $\zeta_p$ is in the limit $r\rightarrow 0$. One might therefore simplify the calculation of the exponents by taking $n$ large and just considering the contributions from the single level $n$. The binomial distribution will then be sharply peaked and we can make use of the normal approximation to the binomial distribution,
\bg
{{n}\choose{k}\, }f_0^kf_1^{n-k}\approx {1\over \sqrt{2\pi n f_0f_1}}\, e^{-{(k-nf_0)^2/ 2nf_0f_1}}\, .
\ee
The peak contribution to $Q$ from structures at level $n$  comes from a scale $r = \lambda_0^{nf_0}\lambda_1^{nf_1}$. 
The approximation then consists of equating the total contribution (\ref{eqQsumnk}) to $Q$, to the value of $Q(r)$ at this dominant scale $r$; in other words we set
\begin{equation}
Q(r) =  (\beta_0 Q_{\lambda_0}+\beta_1Q_{\lambda_1})^n, \quad r = \lambda_0^{nf_0}\lambda_1^{nf_1}. 
\end{equation}
The scaling exponent $\zeta_Q$ linked to $Q(r)$ is then defined by 
\bg
\beta_0 Q_{\lambda_0}+\beta_1Q_{\lambda_1} = \Upsilon^{\zeta_Q},  \quad \Upsilon=\lambda_0^{f_0}\lambda_1^{f_1},
\ela{eqdifflambdas}
\ee
where $f_0,f_1$ are determined by the quantity being averaged through \er{fg}, giving explicitly 
\bg
\zeta_Q = \frac{(\beta_0 Q_{\lambda_0}+\beta_1Q_{\lambda_1} ) \log (\beta_0 Q_{\lambda_0}+\beta_1Q_{\lambda_1} )}
{\beta_0 Q_{\lambda_0}\log \lambda_0 + \beta_1 Q_{\lambda_1}\log \lambda_1 }\, . 
\label{eqdifflambdasexplicit}
\ee

\begin{figure} 
  \centering
  \includegraphics[bb=0 0 420 315,width=3in,height=2.25in,keepaspectratio]{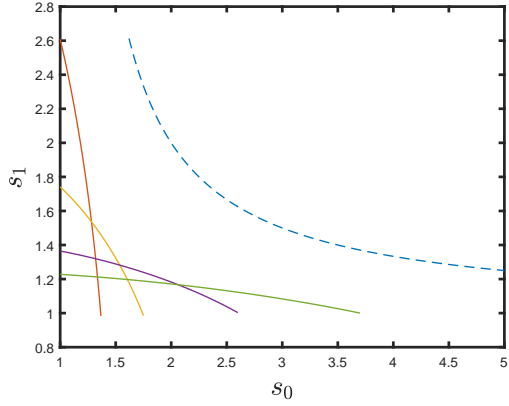}
  \caption{$s_1$ as a function of $s_0$, as determined by solving \er{45law} to satisfy the four-fifths law. The curves, in descending order for their intercepts with the $s_1$ axis, are for $\beta_0=0.7$, $0.5$, $0.3$, $0.2$. The dashed line is the curve $s_1=s_0/(s_0-1)$.}
  \label{fig:fourfiveplot}
\end{figure}

For velocity structure functions, 
$Q_{\lambda_0}= (\beta_0 s_0)^{p/2}$, $Q_{\lambda_1}=(\beta_1s_1)^{p/2}$,  for $p=1,2,3\dots.$, with now $\zeta_Q=\zeta_p$.
In particular the four-fifths law may now be stated as
\bg
\beta_0^{5/2}s_0^{3/2}+\beta_1^{5/2}{s_1}^{3/2}=\Upsilon
=\lambda_0^{\beta_0^{5/2}s_0^{3/2}\over \beta_0^{5/2}s_0^{3/2}+\beta_1^{(5/2)}{s_1}^{3/2}}{\lambda_1}^{\beta_1^{5/2}{s_1}^{3/2}\over \beta_0^{5/2}s_0^{3/2}+\beta_1^{5/2}{s_1}^{3/2}}.
\ela{45law}
\ee
Recall that  $\lambda_0=\sqrt{\beta_0/s_0}$ and $\lambda_1=\sqrt{\beta_1/s_1}$. Thus \er{45law} is a relation between $\beta_0$, $s_0$, $s_1$. In figure \ref{fig:fourfiveplot}
we show this relation. In general to compute structure coefficients we choose $\beta_0$, $s_0$, $\beta_1 = 1- \beta_0$,  find the $s_1$ which enforces the four-fifths law, and then for fixed $\beta_0$ vary $s_0$ until we get good agreement with the She--Leveque formula in a suitable norm; here and below we use the $l^2$ norm, that is the root mean square, but very close results are obtained with the $l^1$ norm. We thus obtain a one-parameter family of ``optimal'' cascades with parameter $\beta_0$.  We have the condition  that $s_0>s_1$ but we also see from figure \ref{fig:fourfiveplot} that the inequality $s_0+s_1 > s_0s_1$ is easily satisfied for the values we use. We shall make use of this inequality in section \ref{time}.  

The above calculation of scaling exponents, yielding (\ref{eqdifflambdasexplicit}), is approximate as the peak contribution to $Q_p(r)$ is not quite at the value of $n$ determined above; however values obtained for key quantities are correct to a few percent. We give a more detailed calculation, which includes the correction to the peak contribution below, in section \ref{sseclargedeviation}. For simplicity, in the tables and figures that follows we use results only from the later, improved calculation.  
 
We next wish to investigate families of models that satisfy the four-fifths law and give structure exponents close to those of She--Leveque in (\ref{eqsl}). We obtain a 1-parameter family of models parameterised by $\beta_0$. To do this systematically we first fix $\beta_0$ and so $\beta_1 = 1 - \beta_0$. We then vary $s_0$ and calculate $s_1$ via imposing the four-fifths law $\zeta_3 = $ (done in (\ref{eqzeta3simples}) below). We then have a model defined by $(\beta_0, \beta_1, s_0, s_1)$: we proceed to calculate the structure exponents $\zeta_p$ for $p = 1, 2, \ldots 10$ (via (\ref{eqldzetap1})) and measure the $l^2$ error for these compared with the She--Leveque ones in (\ref{eqsl}). For a given $\beta_0$ that error is now a function of $s_0$, and we vary $s_0$ to minimise the error. Once we have done this we obtain a branch of solutions parameterised by $\beta_0$.
 
 \begin{figure}[tbp] 
  \centering
(a) \includegraphics[scale=0.28]{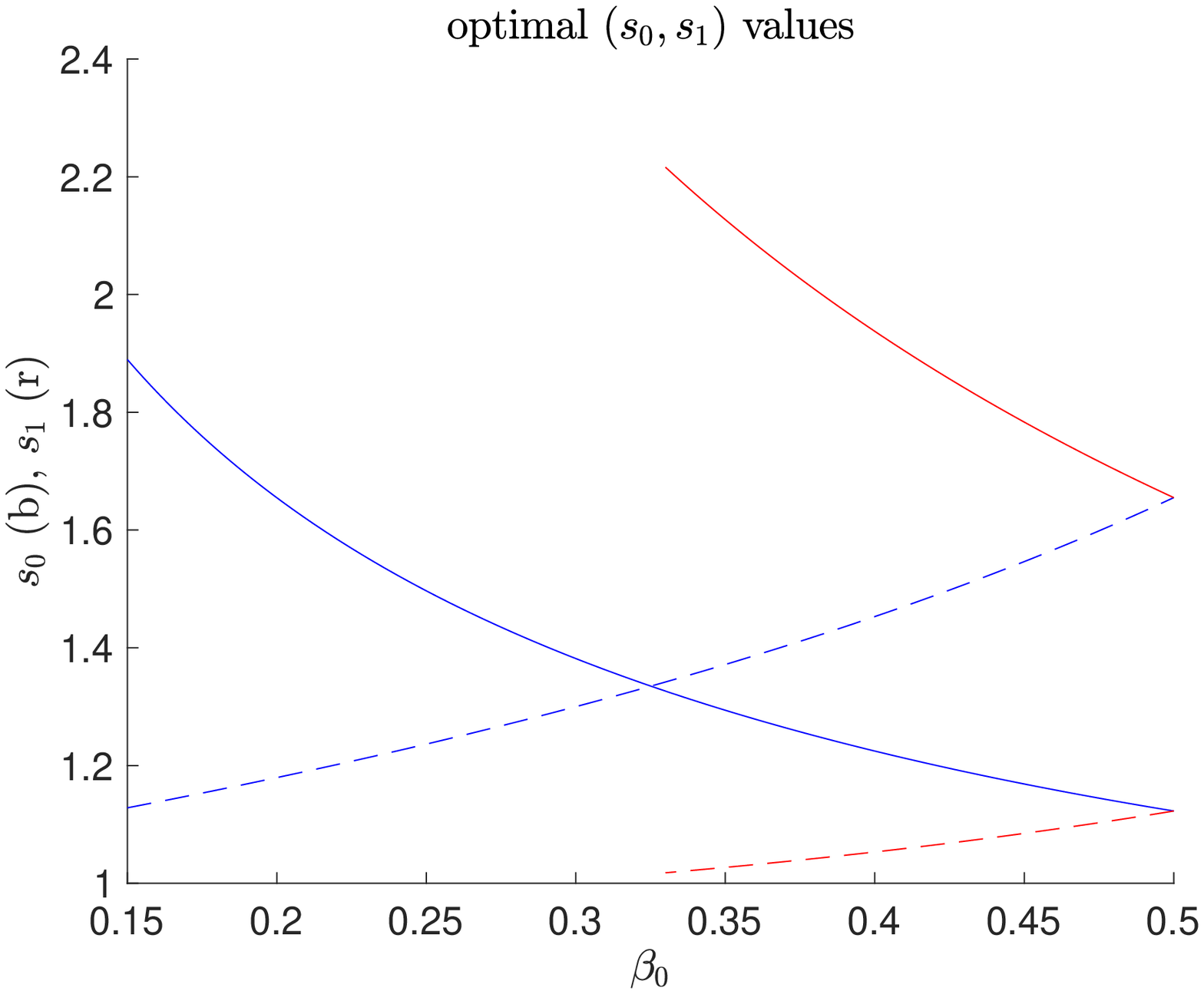}
(b)\includegraphics[scale=0.28]{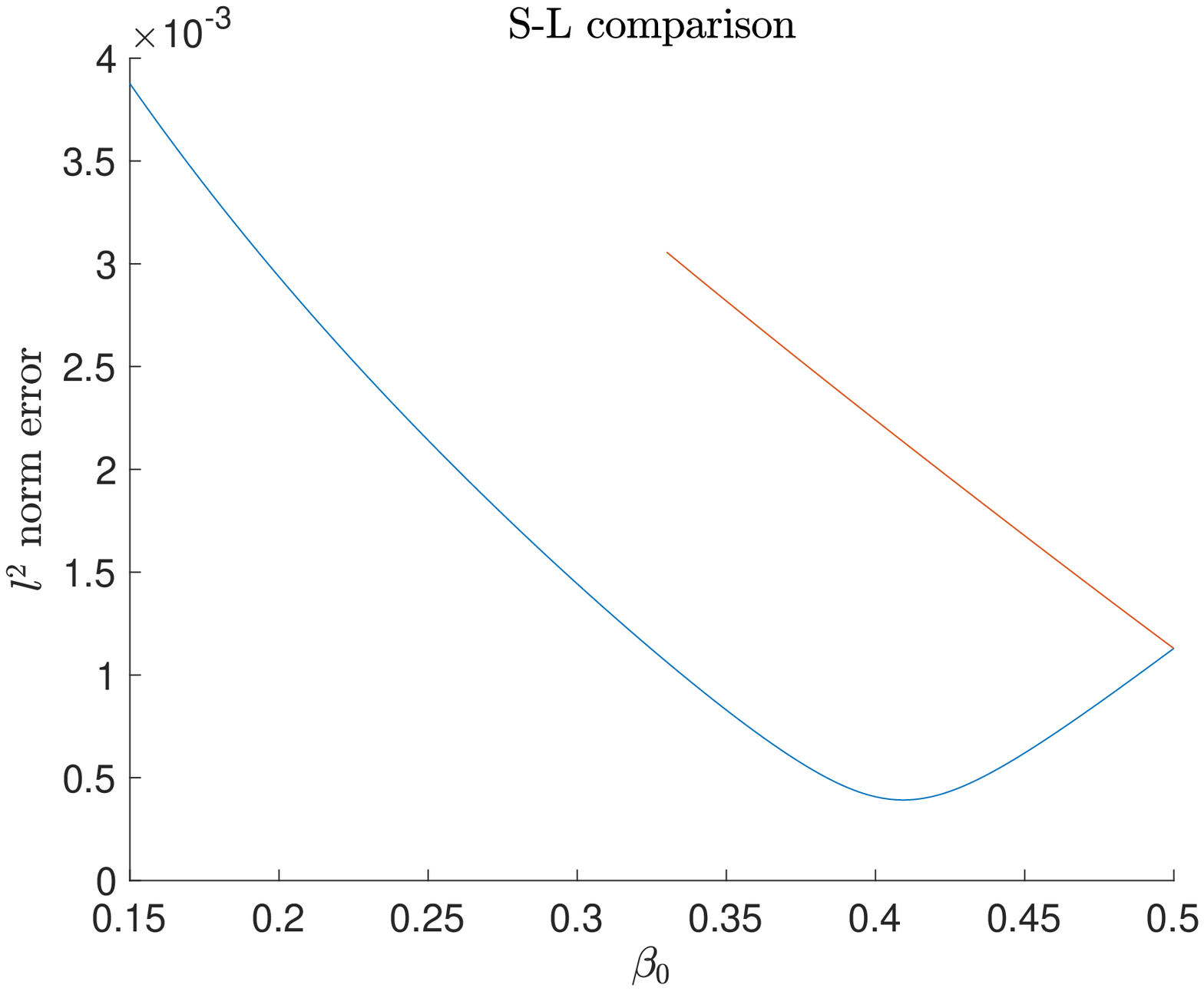}
  \caption{   
 Branches of solutions parameterised by $\beta_0$ that have $\zeta_3=1$ and good agreement with the She--Leveque values of $\zeta_p$. In (a) $s_0$ (solid) and $s_1$ (dashed) are depicted, with blue showing the lower branch and red the upper branch. (b) shows the $l^2$ norm error in the scaling exponents $\zeta_p$ for $p=1, 2, \ldots, 10$, for each branch. }
  \label{fig:adgs0s1norm}
\end{figure}
 
    \begin{table}[h!]
  \begin{center}
    \caption{Calculation of $\zeta_p$ with different values of  $\lambda_(0,1)$ and optimal choice of $\beta_0$, $s_0$, using the large deviation calculation in section \ref{sseclargedeviation}, in particular (\ref{eqldzetap1}). 
    Here $\alpha_i=\log(\sqrt{\beta_(0,1) s_(0,1)})/\log\lambda_(0,1)$ from (\ref{eqalphasdef}) are the scaling exponents for velocity for the cascade branches having subscript $i$. In the table the upper branch is above the lower branch.}
\vskip .1in 
    \label{tab:difflambdas}
    \begin{tabular}{|c|c|c|c|c|c|c|c|c|c|c|} 
    \hline
    $\beta_0$ & $s_0$ & $s_1$ &$\zeta_1$&$\zeta_2$& $\lambda_0$&$\lambda_1$&$\alpha_0$&$\alpha_1$&$s_0/s_1^2$&$s_1/s_0^2$\\
      \hline
0.50 & 1.655 & 1.123 & 0.365 & 0.697 & 0.550 & 0.667 & 0.158 & 0.714 & 1.313&.410\\
0.48 & 1.704 & 1.107 & 0.365 & 0.697 & 0.531 & 0.685 & 0.159 & 0.731 & 1.391&.381\\
0.45 & 1.783 & 1.085 & 0.367 & 0.698 & 0.502 & 0.712 & 0.160 & 0.760 & 1.515&.341\\
0.42 & 1.872 & 1.065 & 0.368 & 0.699 & 0.474 & 0.738 & 0.161 & 0.792 & 1.650&.304\\
0.38 & 2.009 & 1.042 & 0.369 & 0.700 & 0.435 & 0.771 & 0.162 & 0.841 & 1.849&.258\\
0.35 & 2.127 & 1.027 & 0.371 & 0.701 & 0.406 & 0.796 & 0.164 & 0.883 & 2.017&.227\\
0.33 & 2.216 & 1.018 & 0.372 & 0.702 & 0.386 & 0.811 & 0.164 & 0.915 & 2.139&.207\\
\hline
 0.50&    1.123&    1.655&    0.365&    0.697&    0.667&    0.550&    0.714&    0.158&    0.410&    1.313\\
0.45&    1.169&   1.546&    0.363&    0.696&    0.621&    0.596&   0.673&    0.157&    0.489&    1.132\\
0.41&    1.213&    1.471&    0.368&    0.695&    0.582&    0.633&    0.645&    0.155&    0.561&    1.000\\
0.37&    1.264&   1.403  & 0.360  &  0.694&    0.541&   0.670&    0.618&    0.154&    0.642&    0.878\\
0.33 & 1.326 & 1.342 & 0.359 & 0.692 & 0.499 & 0.707 & 0.594 & 0.153 & 0.737&.763\\
0.31 & 1.362 & 1.314 & 0.359 & 0.692 & 0.477 & 0.725 & 0.582 & 0.153 & 0.789&.708\\
0.28 & 1.424 & 1.274 & 0.358 & 0.691 & 0.443 & 0.752 & 0.566 & 0.152 & 0.878&.628\\
0.25 & 1.497 & 1.237 & 0.357 & 0.690 & 0.409 & 0.779 & 0.549 & 0.151 & 0.979&.552\\
0.23 & 1.554 & 1.213 & 0.356 & 0.690 & 0.385 & 0.797 & 0.539 & 0.150 & 1.056&.502\\
0.21 & 1.619 & 1.191 & 0.355 & 0.689 & 0.360 & 0.815 & 0.528 & 0.150 & 1.142&.454\\
0.18 & 1.737 & 1.158 & 0.354 & 0.688 & 0.322 & 0.841 & 0.513 & 0.149 & 1.294&.384\\
\hline
    \end{tabular}
  \end{center}
\end{table}

Our results are shown in table \ref{tab:difflambdas} and figure \ref{fig:adgs0s1norm}, and surprisingly we find two solution branches of acceptable cascade models. Figure \ref{fig:adgs0s1norm}(a) shows values of $s_0$ as solid blue/red curves for the lower/upper branches, as functions of $\beta_0$. The dashed curves show the corresponding values of $s_1$: note that the solution branches are related by the reflection symmetry in the line $\beta_0 = \tfrac{1}{2}$, that is $\beta_0 \leftrightarrow \beta_1$, $s_0 \leftrightarrow s_1$. The fit to the She--Leveque scalings exponents is excellent for both branches, with the $l^2$ (rms) error depicted in figure \ref{fig:adgs0s1norm}(b). The remarkable agreement suggests that our models realise physically the assumptions underlying the She--Leveque result. 

Further analysis of the two branches is given in \ref{tab:difflambdas}. For the lower branch (the values below the middle horizontal line, blue curve in figure \ref{fig:adgs0s1norm}), 
we rejected  larger values of $s_0$ as giving an extremely tight [0] helix ($m[0]/m \approx 16$ when $s_0=3$). For the upper branch  (values above the  line, red curve in figure \ref{fig:adgs0s1norm}), we did not find solutions below $\beta_0\approx 0.33$. Note the last two column, for $s_0/s_1^2$ and $s_1/s_0^2$, eliminate the upper branch if the one  helix is to wrap around the other. On the lower branch, these constraint allow two values a $\beta_0$, one slightly smaller than $0.25$,
the other at $0.41$, with alternate wrapping of the two daughter helices. These values then determine the bimodal cascades of choice in this model.

\subsection{Large deviation calculation of scaling exponents $\zeta_p$} \label{sseclargedeviation}

While this calculation in the previous section leading to the formula (\ref{eqdifflambdasexplicit}) for $\zeta_p$ is useful for exploring the parameter space of cascades, it makes a small error of a few percent in the $s_(0,1)$, though increasing for the smaller $\beta_0$ values on the lower branch. The reason for the error lies in the contributions from multiple levels of the cascade to the eddy size $r$. It might be thought that these would be negligible at large $n$. However the sharp peak in the binomial distribution  forces contributions below the largest $n$ to lie in the tails of the distribution. This requires an application of large deviation theory and use of Stirling's formula in the binomial coefficients. 

We now describe this precise calculation of scaling exponents for large $n$ over a range of cascade levels. 
If we fix a scale $r\ll1$ then contributions to $Q(r)$ will come from a range of levels $n$, those for which $\lambda_0^n \leq r \leq \lambda_1^n$ or $n_- \leq n \leq n_+$, taking $\lambda_0 < \lambda_1$ without loss of generality. For each level $n$ where there is a contribution, this will arise from structures labelled by $k$ with $r\simeq \lambda_0^k \lambda_1^{n-k}$ (or nearby values of $k$), as depicted in figure \ref{fig:knplotadgillus}. Thus, as far as we need for a scaling argument we can write 
\begin{equation}
Q(r) =  \sum_{n=n_-}^{n_+}
 \begin{pmatrix} n\\ k \end{pmatrix} 
(\beta_0 Q_{\lambda_0} )^k (\beta_1 Q_{\lambda_1} )^{n-k}  \bigg|_{k = k(r,n)} 
\end{equation}
with 
\begin{equation}
k (r,n) = \mathrm{round} \left[  \frac{ \log r - n \log \lambda_1}{ \log(\lambda_0 / \lambda_1)}  \right] ; 
\end{equation}
as we vary $n$ we are also varying $k$ to maintain a fixed scale. Setting $b_0 = \beta_0 Q_{\lambda_0}$, $b_1 = \beta_1 Q_{\lambda_1}$ for brevity, we first use Stirling's formula to write 
\begin{align}
& \begin{pmatrix} n\\ k \end{pmatrix} 
b_0^k b_1^{n-k}   =  \frac{1}{ (2 \pi n)^{1/2} (k/n)^{1/2} (1-k/n)^{1/2}} \exp F(n), \\
& F(n)  =   n \log n - k \log k - (n-k) \log(n-k) + k \log b_0 + (n-k) \log b_1 .
\end{align}
We are now able to replace $k$ and $n$ by continuous variables linked by 
\begin{equation}
k (r,n) = \alpha_0 n + \gamma, \quad n-k = \alpha_1 n - \gamma, 
\label{eqklink}
\end{equation}
where
\begin{equation}
\alpha_0 = -  \frac{\log \lambda_1}{\log(\lambda_0 / \lambda_1)}\, , \quad
\alpha_1 = \frac{\log \lambda_0}{\log(\lambda_0 / \lambda_1)}\, , \quad
\gamma = \frac{\log r}{\log ( \lambda_0 / \lambda_1)}\, , 
\end{equation}
noting that $\alpha_0 + \alpha_1 = 1$. With $k$ linked to $n$ via (\ref{eqklink}), we then have that
\begin{equation}
Q(r) =  \int_{-\infty}^{\infty} 
\frac{ \exp F(n)}{ (2 \pi n)^{1/2} (k/n)^{1/2} (1-k/n)^{1/2} }\, dn , \quad k = k(r,n). 
\end{equation}
The contribution is peaked around the maximum of $F(n)$ at say $n = \nb$, in other words where
\begin{align}
F'(\nb) & =    \log \nb   
- \alpha_0 \log (\alpha_0 \nb + \gamma) 
- \alpha_1 \log(\alpha_1 \nb - \gamma) 
+ \alpha_0  \log b_0 + \alpha_1 \log b_1 = 0, 
\end{align}
and we have 
\begin{align}
F(\nb)  
  & =  [ \log ( \lambda_0/\lambda_1)]^{-1}  \bigl[ -    \log [(\alpha_0  + \gamma/\nb)/b_0] +  \log[(\alpha_1 -\gamma/\nb) /b_1] \bigr]   \log r .
\end{align}
In a scaling argument we can ignore algebraic prefactors and focus on the exponential dependence on $F$, to yield
\begin{equation}
Q(r) \sim   \exp F(\nb) \sim r^{\zeta_Q} . 
\end{equation}

To tidy this up, set 
\begin{equation}
\nb / \log r = \delta^{-1}, \quad \ell = \log (\lambda_0/ \lambda_1), 
\end{equation}
and then in a calculation of a scaling exponent we first obtain $\nb$ or equivalently $\delta$, which means solving 
\begin{align}
&  - \log \lambda_1 \log [(\delta - \log \lambda_1) /b_0\ell ]
+ \log \lambda_0  \log[ ( \log \lambda_0  - \delta) /b_1\ell] = 0 
\end{align}
for $\delta$, and then  substituting to obtain 
\begin{align}
& \zeta_Q =  \ell^{-1}  \bigl[ -    \log [(\delta - \log \lambda_1 )/b_0\ell] +  \log[(\log\lambda_0 -\delta) /b_1\ell] \bigr]. 
\end{align}

These two equations are linear in the terms involving the logarithm of $\delta$ and other quantities; these may be solved and then $\delta$ eliminated to leave
\begin{equation}
b_0 \lambda_0^{-\zeta_Q} + b_1 \lambda_1^{-\zeta_Q} = 1 .
\end{equation}
For any choices of $\lambda_(0,1)$ and $b_(0,1) = \beta_(0,1) Q_{\lambda_(0,1)} $ giving the quantity $Q$ we wish to measure, this is the implicit equation for $\zeta_Q$. 

For the case of the $Q_p$ and exponents $\zeta_p$ we have $b_{0,1} = \beta_{0,1}^{p/2 +1} s_{0,1}^{p/2}$, $\lambda_{0,1} = (\beta_{0,1} / s_{0,1})^{1/2} $  and so this becomes 
\begin{equation}
\beta_0^{(p-\zeta_p)/2 + 1}  s_0^{(p+\zeta_p)/2} + \beta_1^{(p-\zeta_p)/2 + 1}  s_1^{(p+\zeta_p)/2} = 1 .
\label{eqldzetap1}
\end{equation}
Imposing $\zeta_3 = 1$ gives a particularly straightforward equation, namely
\begin{equation}
\beta_0^{2}  s_0^{2} + \beta_1^{2}  s_1^{2} = 1 .
\label{eqzeta3simples} 
\end{equation}
giving $s_1$ explicitly in terms of $s_0$ and $\beta_0$ and making $\bar{\varepsilon}_{\mathrm{total}} = 1$.
We thus have established exact constancy of energy flux when $Q_p(r)$ is determined precisely. 
We note that, when $\lambda_0=\lambda_1=\lambda $,  (\ref{eqldzetap1}, \ref{eqzeta3simples}) yield our previous results for this special case. This is not surprising since (\ref{eqldzetap1}, \ref{eqzeta3simples}) are precise for large $n$ and the special case applies to any $n$.

\begin{figure} 
  \centering
  \includegraphics[bb=0 0 420 315,width=3in,height=2.25in,keepaspectratio]{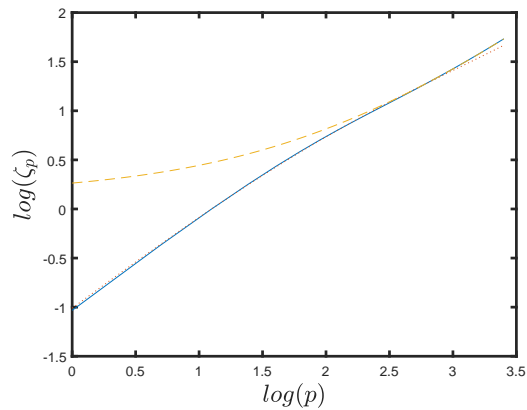}
  \caption{$\log(\zeta_p)$ versus $\log(p)$ for $beta_0=.23$.  The dotted line is the She--Leveque results. The dashed line is the asymptotic approximation \er{asymp}.}
  \label{fig:largep}
\end{figure}

It is of interest to explore the asymptotics of $\zeta_p$ for large $p$ in our model. Then one or other of the terms on the left-hand side of  (\ref{eqldzetap1})  becomes negligible, giving the approximation 
\begin{equation}
\zeta_p \simeq \min_{i=\{0,1\}} ( \tfrac{1}{2} p \log (\beta_i s_i) + \log \beta_i) / \log \lambda_i. 
\ela{asymp}
\end{equation}
For example, for $\beta_0=0.23$ the values in table \ref{tab:difflambdas} the minimum is obtain always for $i=1$. In figure \ref{fig:largep} we compare our values for $\zeta_p$ with She--Leveque and with the above approximation, out to $p=30$.  There is no indication of saturation and our model gives an apparent asymptote somewhat steeper than She--Leveque. It is interesting that it is the parameters with subscript 1 which control the asymptote. This highlights the competition between the two branches of the cascade, giving rise to the nonlinear dependence of $\zeta_p$ upon $p$.

\subsection{Remarks}

We emphasize again the special features of this binary cascade. The two values of  $\alpha_i$, which are the velocity scaling exponents for the two branches of our model, from (\ref{eqalphasdef}), are distinct from K41, although the overall velocity scaling exponent $\zeta_1$ is close to it. We suggest that our model is but one example of a cascade with multiple scalings of velocity. The K41  values in these models would be entirely statistical. It is also interesting that
$\zeta_1$ and $\zeta_2$ tend to lie above K41, a feature that has been observed for low-order structure exponents \cite{sreeni}. Also, although we have two parameters to adjust for agreement with experimental results, the range of permissible values is rather small. If the geometric conditions 
$s_0=s_1^2$ or $s_1=s_0^2$  are imposed, we have seen that this will fix $\beta_0$ as close to either $0.25$ or $0.41$.

One of the tenets of turbulence phenomenology is the ``localness of eddy interactions''. That is, the cascading of eddies of a particular scale is not significantly affected by eddies of much larger or much smaller scale. Our approximate calculation dealt with eddies of a particular ``effective'' scale $\Upsilon$, realized at a  given level $n$. Thus this is a very local calculation, and incurred some error. We propose that the corrections for large deviation can be regarded as an assessment of non-local effects on the cascade. Another approach to non-localness, through the flux of kinetic energy down the cascade, is considered next.

We have not in our model dealt with the {\em helicity} of the cascade. Euler flows conserve total helicity, a measure of the knottedness   of vortex lines \cite{moff},
Contributions to helicity come from both the winding of one filament around another, and from the winding of vortex lines within a filament, the later being associated with axial flow within the filament. Helicity is thus an invariant which is quite sensitive for the dynamics as well as the kinematics of vorticity. We point out that our winding of one filament around another involves a choice of orientation, and this is immaterial to the scaling calculations of this paper. Also we believe these calculations are actually insensitive to  the underlying basic structure, so long as there is stretching and the formation of a succession of self-similar scales. A revisiting of the energy involved would be needed however. An alternative to the helix could be a configuration of rings encircling the parent filament. The rings might alternate in orientation, which is close to the structures observed in \cite{McK, McK2}, see \cite{chilvort}. Other windings which conserve helicity are possible. For these reasons we propose that conservation of helicity is not a determining constraint on the  calculations given in this paper, although it is an essential part of the dynamics.

\subsection{Conservation of energy}
\label{consener}
Kinetic energy is a dynamical quantity, and our model is almost free of dynamical input.  The kinematics of vorticity can yield  accessible vortex structures but cannot yield the energy of a free vortex system moving under self-induction. We have considered in section \ref{energy} approximations relevant to the  energy of a static system of two helical filaments. We can thus compare  the energy of  filament H  with the energy of the split system of two fixed filaments H[0], H[1], including their interaction energy. We shall apply these computations now to see how well energy is conserved.

We make use of \er{totener}, involving $E$, the energy of the unsplit filament, $E[0]$, $E[1]$, the energy of the two daughters, and $E[0,1]$, the interaction energy.
We therefore set $\eee[0]=\beta_0^2s_0E[0]/ E$, $\eee[1]=\beta_1^2s_1 E[1]/E$, $\eee[0,1]=\beta_0\beta_1s_0 E[0,1]/E$. The quantity $\eee_{\rm t}=\eee[0]+\eee[1]+\eee[0,1]$
should therefore be unity if the energy of the split system is the same as that of the unsplit filament.
We show in table \ref{tab:tabener} the results for the cases displayed in table \ref{tab:difflambdas}. 

\begin{table}[h!]
  \begin{center}
    \caption{Calculations relevant to energy conservation of a filament H and daughters H[0] and H[1], with $m=10, \eps=0.01$. The values in table 
\ref{tab:difflambdas} having $s_0>s_1$ are used.}
\vskip .1in 
    \label{tab:tabener}
    \begin{tabular}{|c|c|c|c|c|c|c|c|c|c|c|c|} 
    \hline
    $\beta_0$ & $s_0$ & $s_1$& $b/c$ &$m[0]/m$&$m[1]/m$&$\eee[0]$& $\eee[1]$ & $\eee[0,1] $ & $\eee_{\rm t}$&$\eee_{\rm t}^*$&$\bar{\varepsilon}_\mathrm{total}$\\
      \hline
0.31 & 1.362 & 1.314 &0.274 &    2.855 &    1.813  &  0.152 &   0.682  &  0.414  & 1.247  &  1.039  &  1 \\
0.28 & 1.424 & 1.274 & 0.500  &  3.211 &    1.694&   0.1301&   0.711&   0.354&    1.195&    1.006&    1\\
0.25 & 1.497 & 1.237 & 0.683&    3.663&   1.588&    0.111&    0.740&    0.318&    1.169&    0.995 &   1 \\
0.23 & 1.554 & 1.213 &  0.801&    4.039&    1.522&   0.098&    0.760&    0.298&    1.156&    0.992&    1\\
0.21 & 1.619 & 1.191 &   0.922&    4.495&    1.461&    0.086&    0.780  &  0.279&    1.144&    0.990&    1\\
0.18 & 1.737 & 1.158 & 1.117&    5.396&    1.377&    0.068&    0.810&    0.251&    1.128&    0.990&    1 \\

\hline

    \end{tabular}
  \end{center}
\end{table}

We see that the splitting of the isolated H filament formally requires some energy, since $\eee_{\rm t} > 1$. We have included however another value 
$\eee_{\rm t}^*$ which is in fact less than unity by a small amount. This quantity is an attempt to account for the fact that any filament which splits is part of a cascade, and therefore there will be interaction energy involved. Thus our starting filament H is itself part of a structure of interacting vortical filaments. We will try to account for this fact in the simplest manner, by 
assuming that these interactions are quite local. The idea is to distribute the interaction energy of any pair of daughters, proportionally between the two daughters. As an example consider the case $\beta=0.28$ in table \ref{tab:tabener}. We shall add to $\eee[0]$ a fraction $\eee[0]/(\eee[0]+\eee[1])$ of $E[0,1]$. Then, considered in isolation, H[0] will be assigned an energy  
$\eee[0] + \eee[0,1]\eee[0]/(\eee[0]+\eee[1])$ times $E$, and similarly for $\eee[1]$. 
Thus both $E[0]$ and $E[1]$ are increased by a factor $1+\eee[0,1]/(\eee[0]+\eee[1])$. But the same allocation of interaction energy should apply to all filaments in the cascade, including the one, H, which was our starting filament. Thus the true starting energy was actually $E(1+\eee[0,1]/(\eee[0]+\eee[1]))$, and we should actually multiply the $\eee_{\rm t}$ of table \ref{tab:tabener} by $\eee[0]+\eee[1]$. In this was we obtain  $\eee_{\rm t}^*$. Given the approximations in our calculations and  the ambiguities concerning the dynamics of the cascade  we believe there is a reasonable case to be made for overall energy conservation in our model. We shall see that this ``renormalization'' of energy to account for interactions comes out naturally  when the flow of energy through the cascade is calculated, see section \ref{flow}.

\subsection{Probability density function (pdf) of the velocity}

\begin{figure}[tbp] 
  \centering
  \includegraphics[bb=0 0 420 315,width=2.5in,height=1.87in,keepaspectratio]{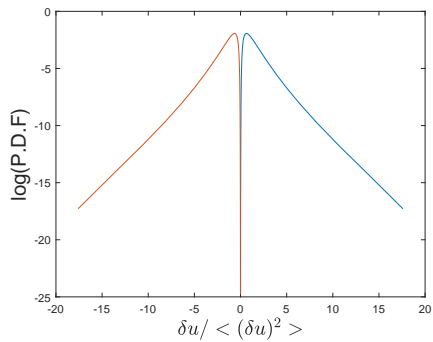}
  \caption{The  velocity pdf  for velocity differences for the case $\lambda_0=\lambda_1$, $\beta_0=0.562$, $s_0=1.52$, $s_1=1.18$, at  or up to   step $N=30$. }
  \label{fig:velpdf}
\end{figure}

It is interesting to see what the pdf of velocity difference $\delta u$ looks like in our model. The simplest way to obtain this is to again take  
$\lambda_0=\lambda_1$, with $\beta=0.562, s_0=1.52, s_1=1.18$, so that all eddy sizes at a given step are the same. We may then compute the pdf for vorticity and normalize to obtain the pdf for velocity differences. We show the result in figure \ref{fig:velpdf}. It has the characteristic non-Gaussian shape of the observations.
The decrease for small values of the velocity is real for our model, and we regard it as a feature of ``inviscid turbulence''. These eddies would normally lie in the dissipation range and have a Gaussian structure.

\section{Timing and the loss of energy}
\label{time}
We turn now to the study of a single cascade as an initial value problem starting from the root filament. Our object is to calculate the flow of energy through our cascade, as well as the time associated with each splitting event. This will allow us to determine the disappearance  of kinetic energy into arbitrarily small spatial scales  in the inviscid limit. We shall say that energy delivered to the smallest scale, which can be arbitrarily small in the inviscid limit, ``dissipates''. The implication is that the energy will be removed by viscous dissipation, the time of delivery of the energy being the time of initiation of dissipation. The time history of the actual process of dissipation is another matter, which we take up briefly at the end of this section.

\subsection{Timing}

In this section times will be in the units of the inverse root vorticity $L/U$. Consider first the times of the cascade steps. We shall focus on the inviscid limit and we shall show that as $n\rightarrow\infty$ every filament has a specific time of formation. These times will be distributed over a finite temporal window where energy is dissipated. This sequence of times will thus determine the decay history of energy in the inviscid limit. 

We assume that the  steps of the cascade take a time inversely proportional to the vorticity of the filament being formed. Let $\tau $ be the time in units of the inverse of the root vorticity, and $\kappa$ the filament vorticity in units of the root vorticity.  
We introduce  the splitting map
\bg
(\kappa,\tau)\rightarrow \Big(s_1\kappa, \tau+{\Lambda \over s_1 \kappa}\Big)\oplus
\Big(s_0\kappa, \tau+{1\over s_0 \kappa}\Big).
\ela{basicmap}
\ee
We have introduced only one time adjustment factor $\Lambda$ since there is an arbitrary unit of time. Introducing $\Lambda_1$, $\Lambda_0$ leads to a calculations involving only $\Lambda_1/\Lambda_0$. This parameter is needed to account for the binary cascade, each splitting producing two different structures. Their interaction could then affect how each filament evolves to the next splitting.

We can also express this map  in terms of operators $\mathcal{O}_{1,0}$ defined by
\bg
\mathcal{O}_{1}(\kappa,\tau)= \Big(s_1\kappa, \tau+{\Lambda\over s_1 \kappa}\Big), \quad 
\mathcal{O}_0(\kappa,\tau)=   \Big(s_0\kappa, \tau+{1\over s_0 \kappa}\Big).
\ee
These are commuting operators if and only if  $\Lambda(s_0-1)=(s_1-1)$. Thus $n$ steps of the cascade can be represented by 
\bg
(\mathcal{O}_1\oplus \mathcal{O}_{0})^n, 
\ee
as an ordered product.

 We summarize now the various orderings satisfied by $s_0,s_1$, as they will be needed below:
\bg
  s_1>1, \quad s_0>1,\quad {s_1-1\over s_0-1} < {s_1\over s_0}< 1,\quad  s_0+s_1-s_0s_1 > 0.
\ela{orderings}
\ee 
The first two express our assumption that thin filaments will always be stretched. The middle inequalities follow from the need to stretch the [0] helix in order that it wrap around the [1] helix. The last inequality expresses results of the computations within our model: it is found to be easily satisfied for the parameter values used, as indicated in figure \ref{fig:fourfiveplot}.

Referring to figure \ref{fig:Bigtree}, we start at the root with $(\kappa,\tau)=(1,0)$. Then H[1] corresponds to state $ \mco_1(1,0)=(s_1,\Lambda/s_1)$ and H[01] to
$(s_1s_0,1/s_0+\Lambda/(s_1s_0)$. Time ordering will be indicated by $\succ$ or $\prec$. Thus H[11] $\succ$ H[10] if and only if
$\Lambda/s_1+\Lambda/s_1^2 \geq \Lambda/s_1+1/(s_0s_1)$, which here amounts to $\Lambda/s_1 \geq 1/s_0$. In fact if the latter ordering holds then at any step we will have H[$b$1] $\succ$ H[$b$0] for any binary $b$, i.e.\ it holds for all pairs of filaments with a common parent. What about other adjacent filaments? 
For example H[10] $\succ$ H[01] iff $\Lambda/s_1+1/(s_0s_1)\geq 1/s_0+\Lambda/(s_1s_0)$, or $(s_1-1)\leq \Lambda(s_0-1)$. We thus see the relevance of the inequalities in \er{orderings}

Turning to the  the inviscid limit, consider the filaments H[$\overline{0}$] and H[$\overline{1}$], where  the line over a binary sequence means continuation periodically of the digit(s) beneath. We have
\bg
{\rm H[}\overline{0}{{\rm{]}}}={1\over s_0}+{1\over s_0^2}+\dots = {1\over s_0-1}\, , \quad 
 {\rm H[}\overline{1}{{\rm ]}}={\Lambda\over s_1-1}\, . 
\ee
The ordered pair  H[$\overline{1}$], H[$\overline{0}$] determines what we call the {\it root branch}. A \emph{branch} B($b$) will consist of an ordered pair  of the form H[$b\overline{1}$], H[$b\overline{0}$].  Thus a branch is an inverted V with vertex determined by the binary numbers $b$. As we progress down the tree on B($b$) we can define each subsequent  V as a sub-branch SB($b$) of B($b$). 
  
The \emph{time span} of the branch will be the difference in times between two limbs extended to infinity. We write this as $\big[$H[$b\overline{1}$], H[$b\overline{0}$]$\big]$. Thus
\bg
\big[{\rm H[}\overline{1}{\rm ]}, {\rm H[}\overline{0}{\rm ]}\big]= {\Lambda\over s_1-1}-{1\over s_0-1} \equiv {}T
\ee
is the time span of the root branch. 

\begin{figure} 
  \centering
  \includegraphics[bb=0 0 552 634,width=3in,height=3.45in,keepaspectratio]{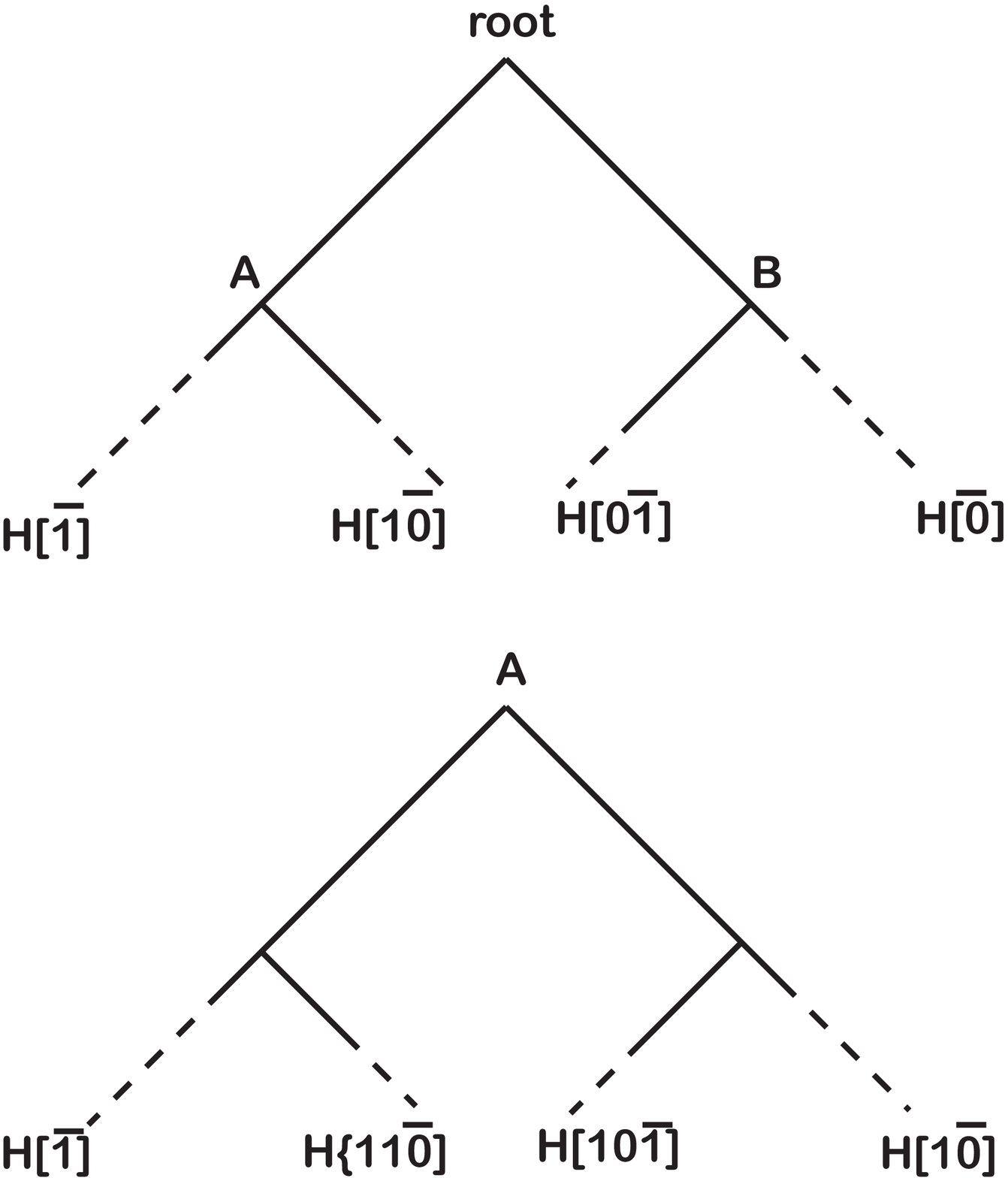}
  \caption{A representation of filament times. Here the dashed extensions indicate continuation to infinite steps. The lower tree is a branch of the upper one. }
  \label{fig:dottree}
\end{figure}

Referring to figure \ref{fig:dottree},  consider the sub-branches of the root $\mathrm{SB}[0] = (\HH[0\overline{1}], \HH[\overline{0} ])$  and $\mathrm{SB}[1] = (\HH[\overline{1}], \HH[1\overline{0}])$. We have
\bg
 \big[{\rm{H[}}\overline{1}{{\rm ]}},{{\rm H[}}1\overline{0}{{\rm ]}}\big]={\Lambda\over  s_1-1}-{\Lambda\over s_1}-{1 \over s_1(s_0-1)}={{\it T }\over s_1}\, .
\ee
Similarly
\bg
\big[{\rm{H[}}0\overline{1}{\rm ]},{\rm H[}\overline{0}{\rm ]}\big]={{\it T }\over s_0}\, .
\ee
Carrying this one step further we find
\bg
\big[{\rm{H[}}\overline{1}{\rm ]},{\rm H[}11\overline{0}{\rm ]}\big]={{\it T }\over s_1^2}\, , \quad
\big[{\rm{H[}}10\overline{1}{\rm ]},{\rm H[}1\overline{0}{\rm ]}\big]={{\it T }\over s_1s_0}\, .
\ee
These are the two sub-branches  SB[11] and SB[10] emanating from point A in figure \ref{fig:dottree}. From point B of the figure (not the root branch label)
we similarly have,  for the sub-branches SB[01], SB[00]
\bg
\big[{{\rm{H[}}}0\overline{1}{\rm ]},{\rm H[}01\overline{0}{\rm ]}\big]={{\it T }\over s_0 s_1}\, ,  \quad
\big[{{\rm{H[}}}00\overline{1}{\rm ]},{\rm H[}\overline{0}{\rm ]}\big]={{\it T }\over s_0^2}\, .
\ee

The pattern should be now clear. A binary distribution of factors in $s_0^{-1},s_1^{-1}$ of the time span of the root branch apply to the sub-branches. Since at every step, each filament terminates at one end of a branch, we can determine the sequence of times provided we know the temporal separation of branches.

We thus introduce {\it gaps}. The \emph{root gap} is the ordered pair $\GG = (\HH[1\overline{0}], \HH[0\overline{1}])$.   We see that this pair consists
of the right limb of SB$[1]$ and the left limb of SB$[0]$. The gap referred to is the gap between these two sub-branches, .
Proceeding down the tree, in general a sub-gap SG[b]  consists of an ordered pair of the form $ \GG[b] = ( \HH[b1\overline{0}], \HH[b0\overline{1}]$).

We are interested in the time interval associated with a gap. 
The time interval of the root gap is
\bg
\big[{{\rm{H[}}}1\overline{0}{\rm ]},{\rm H[}0\overline{1}{\rm ]}\big]={\Lambda\over s_1}+{1\over s_1(s_0-1)}-{1\over s_0}-{\Lambda\over s_0(s_1-1)}=-{DT\over s_1s_0}\, ,
\ee
where $D=s_0+s_1-s_0s_1 > 0$. We now note that a sequence of binomial factors in $s_0^{-1},s_1^{-1}$ will apply also to gaps. For example the sub-gaps SG[1] and SG[0] have time intervals
\bg
\big[{\rm{H[}}11\overline{0}{\rm ]},{\rm H[}10\overline{1}{\rm ]}\big] =-{DT\over s_1^2s_0}\, , \quad
\big[{\rm{H[}}01\overline{0}{\rm ]},{\rm H[}00\overline{1}{\rm ]}\big] =-{DT\over s_1s_0^2}\, .
\ee
%
Thus the time spans of adjacent branches overlap, irrespective of the sign of $T$. For example the time spans of B$[1]$ and B$[0]$ sum to $T({1/ s_1}+{1/ s_0})>T$ with gap $T({1 / s_1}+{1/s_0}-1)$.
%

If  $\Lambda > {(s_1-1)/( s_0-1)}$ then the first filament dissipated is the last [0] filament, which seems reasonable physically. But the value of $\Lambda$ cannot be determined within our model. If $\Lambda < (s_1-1)/ (s_0-1)$
the last [1] filament formed dissipates first.

\subsection{The commutative case ${\Lambda}=(s_1-1)/(s_0 -1)$}

We now show that for this limit case all time spans shrink to zero and all filaments terminate together. To see this, select a large value of $n$ and select any path through the cascade, up to level $n$.  Because $\mco_1$ and $\mco_0$ commute, we can compute the time from the operator $\mco_0^k\mco_1^{n-k}$ for some 
$k$, $0 \leq k \leq n$. The time so computed is then  given by
\bg
\tau_n\equiv {\Lambda\over s_1-1}\,  (1-s_1^{-(n-k)})+{1\over s_0-1} \, s_1^{-(n-k)} (1-s_0^{-k}) ={\Lambda\over s_1-1} \, (1-s_1^{-(n-k)}s_0^{-k}).
\ee
Then, since $s_0> s_1$
\bg
\Big|\tau_n-{\Lambda\over s_1-1}\Big| < {\Lambda\over s_1-1}\, s_1^{-n}\rightarrow 0,\;\;n\rightarrow\infty.
\ee
Thus all filaments terminate at time ${\Lambda/(s_1-1)}={1/ ( s_0-1)}$. The dissipation history is thus instantaneous. Of course this is improbable as a realistic Euler limit because it depends so much on our highly structured and carefully scaled cascade.

\subsection{The flow of energy}
\label{flow}
We now introduce the other element we must follow, namely the energy within the cascade of helical filaments. This study is complicated by the fact that a given branch of our tree must be considered with the neighboring branch, the two making up a pair of filaments which interact in the energy. We want to replace this situation by an energy associated with a single state, which divides up the interaction energy, along the lines already described in section \ref{model}. Again we often use ``energy'' when we mean ``energy factor''. As an example we shall give specific numbers for the case $\beta_0=0.28$, $s_0=1.5$, $s_1=1.2735$. From the root filament H$[0]$ receives  roughly $\eee[0]=0.14$ of the energy, H$[1]$ receives  about $\eee[1]=0.71$. We will now however deal with the first step of the cascade and assume  energy conservation. That is, we disregard for this step the slight energy excess we calculated for the splitting of an isolated filament. Thus at this stage the interaction energy interaction energy is  $1-\eee[0]-\eee[1]=0.15$.  Distributing this as before, we associate  energy $\eee[0](1+(1-\eee[0]-\eee[1])/(\eee[0]+\eee[1])= \eee[0]/(\eee[0]+\eee[1])$ with H$[0]$ and $\eee[1]/(\eee[0]+\eee[1]) $ with H$[1]$. 
The denominator here is precisely the amplification factor or `renormalization' that was invoked in section \ref{consener}, although there we did not have exact conservation of energy in the splitting.

Consider now H$[11]$ and H$[10]$, with energies ${\rm  e[1]}^2$ and  $\eee[1]\eee[0]$ when not interacting. Their pairwise interaction would then yield, by the same division of interaction energy and addition of part to  ${\rm  e[1]}^2$, giving
\bg
{\rm  e[1]}^2+\bigg({{\rm  e[1]}^2\over {\rm  e[1]}^2+{\rm  e[1]e[0]}}\bigg) ({\rm  e[1]}-{\rm  e[1]}^2-{\rm  e[1]e[0]})= {{\rm  e[1]}^2\over {\rm e[0]+e[1]}}\, .
\ee
Similarly the addition of interaction energies to H$[10]$, H$[01]$, H$[00]$ results in division of the bare energies by $\eee[0]+\eee[1]$. If we now sum these four modified energies we get not unity but rather $\eee[0]+\eee[1]$. Thus conservation of energy in going from step 1 to step two requires a modification that should reflect interaction between the two pairs of filaments. We take this to be the same for all four filaments and so divide the modified energies by $\eee[0]+\eee[1]$, the renormalized energy for H$[11]$ now being
\bg
{{\rm  e[1]}^2\over {\rm (e[0]+e[1])}^2}\, .
\ee
and similarly  for the other three. 

\begin{figure} 
  \centering
  \includegraphics[bb=0 0 420 315,width=3in,height=2.25in,keepaspectratio]{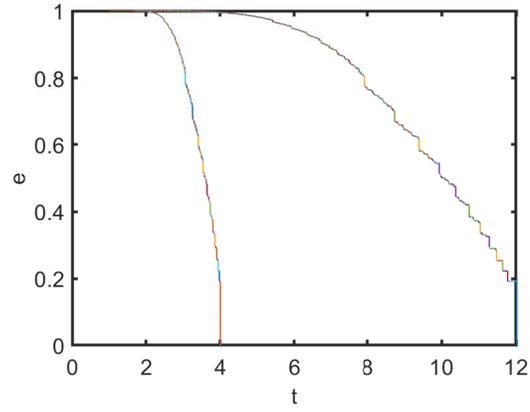}
  \caption{Energy factor e versus time with dissipation at step $N=11$. Here $s_0=1.6$, $s_1=1.2$,  $e[0]=0.14$, $e[1]=0.71$. The left curve is for $\Lambda=3$, the right for $\Lambda=1$. }
  \label{fig:energydecay}
\end{figure}

\begin{figure}
\begin{center}
\includegraphics[scale=0.4]{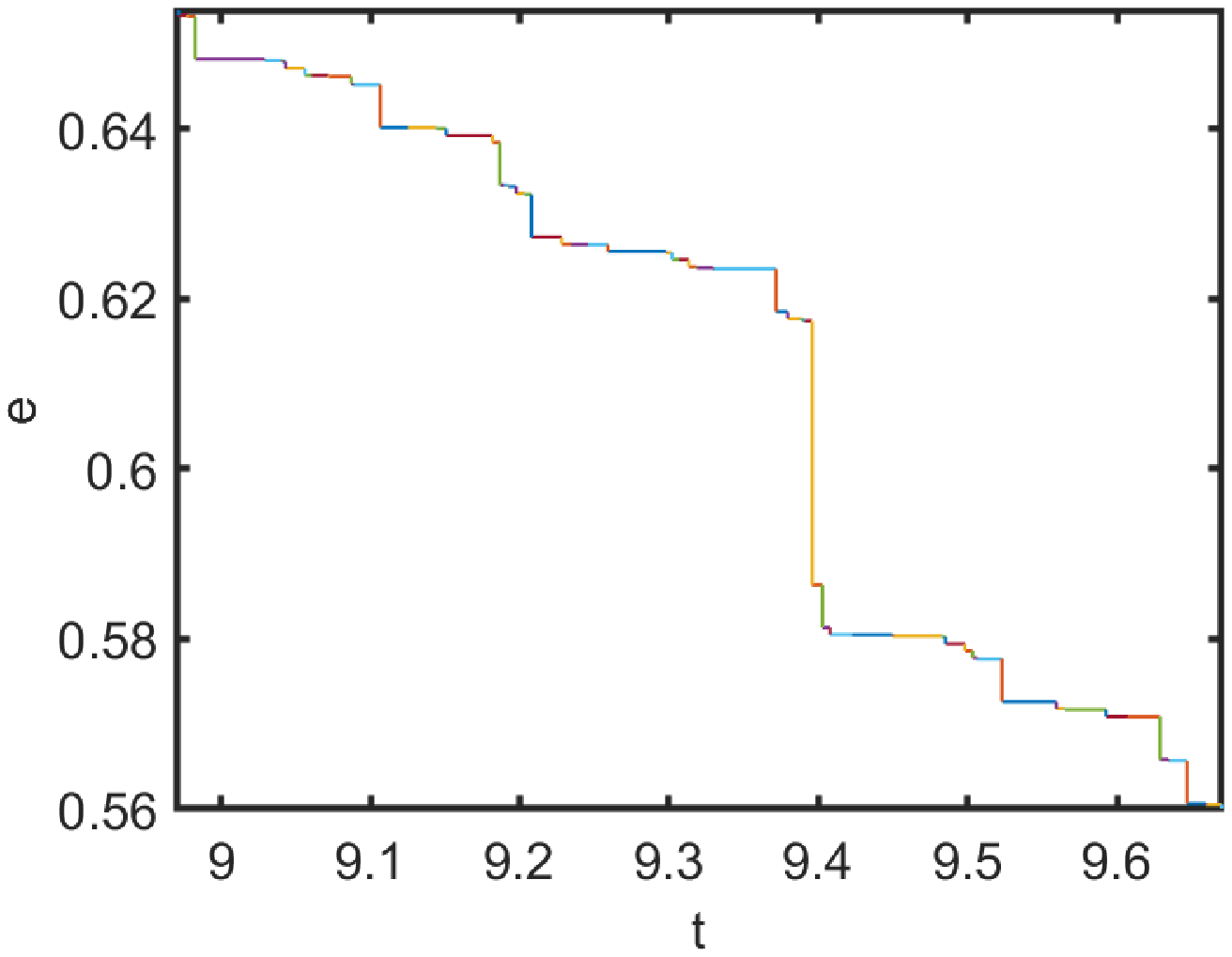} 
\includegraphics[scale=0.4]{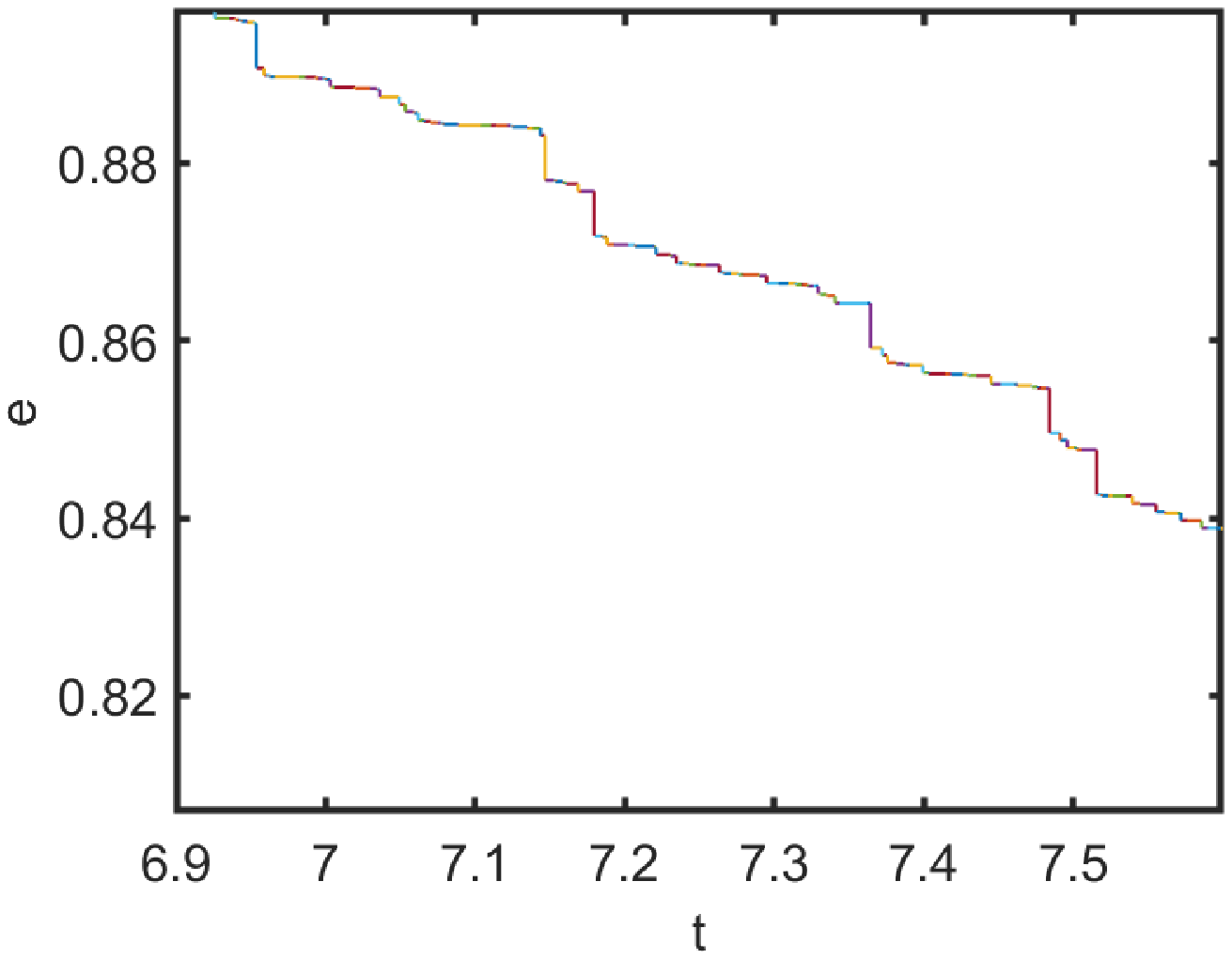}  
\end{center}
\caption{Two small time windows from the case $\Lambda=3$ in figure \ref{fig:energydecay}.}
\label{fig:windows}
\end{figure}

This renormalization is the same at each step, leading to the binomial distribution of energies at step $n$:
\bg
{\rm (e[0]+e[1])}^{-n} {{n}\choose{k}}{\rm e[0]}^{k}{\rm e[1]}^{n-k}.
\ee
We show in figure \ref{fig:energydecay} plots of the resulting dissipation history. We follow levels $n$ for $0\leq n \leq N$  with $N=10$ and assume that energy disappears at the $N$th step. There is at any finite $N$ a slight error since all eddies should have a common  size (physically, the Kolmogorov length). Here we are really taking $N=10$ so all eddies are small but not exactly the same. In this inviscid limit the error disappears. In the figure the curves are for $\Lambda=1$, $3$, and in figure \ref{fig:windows} we show the short time windows for the case $\Lambda=3$. Note the lack of self-similarity: the dissipation history resembles a devil's staircase, often cited as an example  of intermittency, as discussed  in for example \cite{frisch}, p.\ 123.

\subsection{Inviscid decay of energy}
We have calculated the time sequence of delivery of energy to arbitrarily small scales. The question remains of how energy will actually decay under the actidon of (small) viscosity. One of the remarkable properties of a straight vortex filament carrying circulation $\Gamma$ can be seen by considering the exact solution
\bg
 \omega(r,t) = {\Gamma\over 4\pi \nu t} \, e^{-{r^2/ 4 \nu t}}
\ee
of f the 2D Navier-Stokes equations. The resulting dissipation is  
${\Gamma^2/ 8\pi t}$ per unit length, and is independent of viscosity. When multiplied by a length this becomes energy over time. This result might apply locally in space and time to an arbitrary closed filament, giving the initial rate of dissipation, but then nonlinear effects come into play, the filament evolves, reconnection occurs, and a complex dynamic is needed to determine the decay of energy. One possibility is that filamentary modelling continues to apply but that reconnection changes the  energy, so that the decay of total energy $E(t)$ is described by
\bg
 {dE\over dt} = -C\, {E\over t}.
\ee
Then $E$ would decay as $1/t^C$. Kolmogorov calculated $C$ as 10/7, and Saffman has proposed a value of 1.2 \cite{saff}. An estimate of $C$ would appear to lie outside the scope of the present model.

\subsection{Filament geometry}

\begin{figure} 
  \centering
  \includegraphics[bb=0 0 581 613,width=3in,height=3.17in,keepaspectratio]{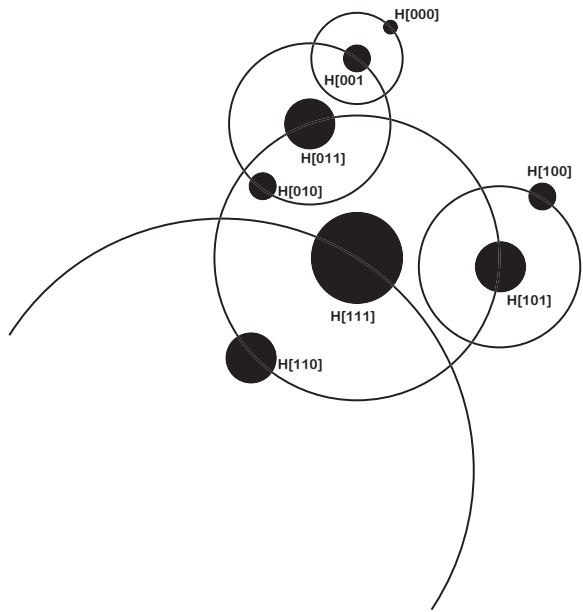}
  \caption{Helical filaments generated by three steps from a root filament assuming that the smaller filament is wound around the larger one.}
  \label{fig:lastplacing3}
\end{figure}

We have exploited  helical geometry as a convenient way to visualize the binary self-similarity of the cascade. We have also made explicit use of it in the calculation of energy and the flow of energy down the cascade. However we must recognize that once filaments are created, even if almost helical, they will rapidly undergo distortion. Thus insofar as placement of the helical filaments is concerned, it only makes sense to consider a few steps of the cascade. We have seen that our cascade is highly localized in that energies need only  be calculated  for a small few steps of the cascade. Here we shall assume that only three steps are needed to  account adequately for the energy of interaction between filaments. With that assumption we can view each splitting as the winding of an H[$b$0] filament around
an H[$b$1] filament. If we take $\lambda_0/\lambda_1\approx 0.5$, which roughly corresponds to case $\beta_0=0.25$ of table \ref{tab:difflambdas}, the resulting structure is depicted in figure \ref{fig:lastplacing3}. We are showing the intersection with a  plane passing through the cores the eight descendants of a single filament, the effective root,  after three steps of the cascade. The circles represent the surface of the tori on which the filaments are wound. Any two filaments having the same first and second  digits, but differing in their third digit, represent a splitting event and are closely interacting. 
We can think of this sketch as an attempt to find order in a cascade of vorticity by focusing on a short time window and a small Lagrangian domain moving with the root filament.

\section{Discussion}

The present study was inspired by the beautiful experimental results reported recently by McKeown {\it et al.} \cite{McK,McK2}. In these experiments and related simulations vortices generated by colliding vortex rings are found to undergo an elliptical instability, leading to the formation of smaller structures, presumably capable again, through their interaction,  of another instability.
This suggested that the models considered in \cite{chilvort}, involving vortex rings upon rings, or helices upon helices, might be worthy of further study. In that paper the emphasis was on modeling a fractal beta model \cite{frisch}.  It became of interest to widen the consideration to a model which accounts in some way for all of the vorticity, thus preserving vorticity volume.

The model proposed here does not pretend to be more than a toy model of the inertial range. Indeed the mechanism observed in \cite{McK} for the production of daughter vortices from a parent  involves the pulling out of a hairpin structure from the surface of the vortex, forming perpendicular daughter rings encircling the parent. The elliptical instability causing these rings is on the scale of the vortex core. This is somewhat reminiscent of  ``rings on rings'', but quite different from  the splitting of thin helices which we invoke here. Nevertheless the structure we propose does bear some resemblance to shredding of unstable vortices seen 
by McKeown {et al.} \cite{McK}

The physical structure considered here can reproduce with remarkable accuracy the She--Leveque expression for $\zeta_p$. For example, with $\lambda_0=\lambda_1$ and $\beta_0=0.4376$ the ratio of the two expressions varies from 0.9954 to 1.0037  for $p$ to 10. Dubrulle and others have shown that the She--Leveque curve corresponds to a log-Poisson distribution; see \cite{frisch}, sec. 8.9.2, and  \cite{dub}. Our model may thus be viewed as a particular realization of this distribution through the splitting and stretching of vorticity.

Our model does have  some features of which may be applied more generally. We have shown that the presence of two distinct scaling factors leads to a workable model of the inertial range which is quite different from the Richardson--Kolmogorov cascade, with $\delta u/U \sim (r / L)^{1/3}$.  While it is to be expected that a model with so few global Eulerian constraints will have free parameters, it is surprising  that we cannot vary $\beta_0$ and $s_0$ that much and still achieve good agreement with experiment as well as realistic splitting geometry. This bodes well for the possibility of more realistic vortical cascades of the inertial range.

The geometry of our model has some appealing features. The structure is built up with minimal surgery on the vorticity. It is true that splitting of vortex tubes by a smooth velocity field will necessarily  produce thin sheets connecting the vortex tubes and these sheets are neglected here. But cutting across a vortex tube is avoided in our model. Such cutting and the accompanying pasting must be expelled from an essentially inviscid inertial range, although viscous reconnection very likely plays an important role in the dissipation range of stationary turbulence and in the decay of free vorticity at large Reynolds numbers.

The fact that the model involves two distinct scalings suggests that there might be some connection to the multi-fractal models of  velocity intermittency \cite{frisch}, since in principle any $\zeta_p$ can be so represented. However there is as yet no physical description of the mechanism of the multi-fractals. The bifractal model involves a piecewise linear $\zeta_p(p)$ and essentially pieces together two beta models, which are operative over adjacent intervals of $p$ (see  \cite{frisch}, section 8.5.2). Our branching model blends two cascades seamlessly and leads easily to the non-linear dependence of $\zeta_p$  upon $p$ associated with intermittency.

\appendix
\section{Calculation of energy for a helical filament}

For our calculation of energy $E$ of an isolated closed helical filament we start with \er{enercalc}. We can think of this integral as that part giving the sum of $\tfrac{1}{2} |\uv|^2$ integrated over the exterior of the filament, and that giving the sum of  $\tfrac{1}{2} |\uv|^2$ over the interior of the filament, the exterior and interior energies. Now the exterior energy depends only on the circulation, not the distribution of vorticity over the core. If we concentrate the vorticity at the edge of the tube, then
there is no internal velocity and \er{enercalc} yields the external energy. But if this is computed from velocity with $\uv=\bmnab\times\Av$ we see by the divergence theorem, assuming sufficent fall-off of $\bmom$ at infinity, and for the concentrated vorticity assumed, that
\bg
E_{\rm ext} = {1\over 2} \int \Av\cdot\bmom \, dV= {\Gamma\over 2} \oint_{\rm axis} \Av_{\rm boundary}\cdot d\Rv, 
\ee
since for slender tubes $\Av$ is approximately constant on the tube boundary.  Also we have
\bg
\Av\approx {\Gamma\over 4\pi}\oint_{\rm axis} {d\Rv^\prime\over |\Rv-\Rv^\prime|}\, , 
\ee 
and make a standard regularization of the singularity for a thin filament  vortex with a circular core. Consider a  integral over a  closed filament.
\bg
\Iv=\oint {d\Rv\over |\Rv_0-\Rv|}\,  ,
\ee  
where $\Rv_0$ is a point of the filament axis. 
Let $r$ be the radius of the core. We choose a scale $\Delta$ large with respect to $r$ but small compared to the filament length. We then divide the integral into two parts:
\bg
\Iv=\oint_{|\Rv_0-\Rv|>\Delta} { d\Rv \over |\Rv_0-\Rv|}  + \int_{|\Rv_0-\Rv|\leq\Delta} {d\Rv\over |\Rv_0-\Rv|}   \, .
\ela{Iint}
\ee
The first integral, $\Iv_1$, may be reduced to a line integral of the circulation with respect to arc length around the axis. The second integral, $\Iv_2$, is over a small thin cylinder of length $2\Delta$ and diameter $2r$. If $\tv_0$ is the tangent vector at $\Rv_0$ we find
\bg
\Iv_2\approx 2\tv_0\log \frac{2\Delta}{r}=\int_{r/2 \leq |\Rv_0-\Rv| \leq \Delta}{d\Rv\over |\Rv_0-\Rv|}\, ,\quad \Delta \gg r . 
\ee
Using this in \er{Iint} we obtain the regularization
\bg
\Iv=\oint_{{\rm axis},\, |\Rv_0-\Rv|>r/2} \; {  d\Rv \over |\Rv_0-\Rv|}\, .
\ee
Thus
\bg
\Av_{\rm boundary}\approx {\Gamma\over 4\pi}\oint_{{\rm axis},\, |\Rv-\Rv^\prime|>r/2}  \;    {d\Rv^\prime\over |\Rv-\Rv^\prime|}\, ,
\ee 
and 
\bg
E_{\rm ext} \approx {\Gamma^2\over 8\pi} \oint_{\rm axis}\oint_{{\rm axis},\, |\Rv-\Rv^\prime|>r/2}\;   {d\Rv \cdot d\Rv^\prime\over |\Rv-\Rv^\prime|}\, .
\ela{enerextfin}
\ee

We now apply this to the helical winding of interest to our model. We first do a computation of energy for a helical filament of turn radius $b$ wound around a large torus of radius $R=mc$. The equation for the points of the filament in Cartesian coordinates is
\bg
 \Rv(t)=((mc+b\cos t)\cos(t/m), (mc + b \cos t)\sin(t/m), b\sin t),\quad 0 \leq t\leq 2\pi m.
\ee
Thus there will be $m$ turns of the helix on the torus. We are interested in the limit of   $E/m$ for large $m$. The exterior energy is given, according to \er{enerextfin}, by
\bg
E_{\rm ext}= {\Gamma^2\over 4\pi }\int_0^{2\pi m}\int_{t^\prime+\eps}^{t^\prime+\pi m } {\tv\cdot\tv^\prime\over |\Rv(t)-\Rv(t^\prime)|} \, dt \, dt^\prime,
\ee
where $\eps= \tfrac{1}{2} {r/ \sqrt{b^2+c^2}}$.  
Note that here $\tv=d\Rv/dt$ and is not the unit tangent vector.
After some calculation we find
\begin{align}
 |\Rv(t)-& \Rv(t^\prime)|^2 = 4b^2\sin^2((t-t^\prime)/2)
+4(mc+b\cos t)(mc+b\cos t^\prime)\sin^2[(t-t^\prime)/2m],
\\
\tv\cdot\tv^\prime & =b^2\big[\sin t \sin t^\prime  \cos [(t-t^\prime)/m]+\cos t \cos t^\prime\big] + c^2 \cos[(t-t^\prime)/m]
\notag\\
& + bc\big[ \sin(t/m)\sin t^\prime \cos(t^\prime/m)+\sin(t^\prime/m)\sin t\cos(t/m)
\label{eqmesstvtv}
\\
& -\cos(t^\prime/m)\sin t\sin (t/m)-\cos(t/m)\sin t^\prime \sin (t^\prime/m)\big]+ O(1/m).
\notag
\end{align}
The estimate is uniform in $t$, $t^\prime$.

Let us first check that we get the right result when $b=0$. Then we have
\bg
E_{\rm ext}= {\Gamma^2 cm\over 4}\int_{r/ 2 mc}^\pi {\cos\psi\over \sin \tfrac{1}{2} \psi }\, d\psi,
\ee
which is precisely the exterior energy of a ring filament of radius $R=mc$.

Now consider the full problem for large $m$. The $mc$ terms of $|\Rv(t)-\Rv(t^\prime)|$ make this $O(m)$ unless $(t-t^\prime)/m$ is 
small. We may divide up the integral into two parts. First, assume that in the $t$ integral $t-t^\prime$ is less than $Am^\alpha$ for some $0 < \alpha < 1$ that we can specify later. Then
\[
\tv\cdot\tv^\prime\approx b^2\big[\sin t \sin t^\prime  +\cos t \cos t^\prime\big] + c^2 
+ bc\big[ \sin(t^\prime/m)\sin t^\prime \cos(t^\prime/m)+\sin(t^\prime/m)\sin t\cos(t^\prime/m)
\]
\bg
-\cos(t^\prime/m)\sin t\sin (t^\prime/m)-\cos(t^\prime/m)\sin t^\prime \sin (t^\prime/m)\big]=b^2\cos(t-t^\prime)+c^2.
\ee 
Thus we find the \emph{inner} contribution
\bg
E_{\rm ext}^{\rm in}\approx {\Gamma^2m\over 2}\int_\eps^{Am^\alpha}{b^2\cos\psi+c^2\over ({4b^2\sin^2\tfrac{1}{2} \psi +c^2\psi^2})^{1/2}} \, d\psi.
\ee

For the \emph{outer} contribution we use the approximation 
$|\Rv(t)-\Rv(t^\prime)|\approx 2cm|\sin[(t-t^\prime)/ 2m]|$. Also we write $\tv\cdot\tv^\prime = c^2\cos[(t-t^\prime)/m]+B(t,t^\prime)$ for (\ref{eqmesstvtv}).
Now
\bg
\int_0^{2m\pi} B(t^\prime+m\psi,t^\prime)\ dt^\prime= b^2\pi m \cos(m\psi)(\cos \psi + 1),
\ee
for fixed $\psi$.  
Then we have
\bg
E_{\rm ext}^{\rm out} \approx {\Gamma^2\over 4\pi} \int_{Am^{\alpha -1}}^\pi { 2\pi m c^2\cos\psi+b^2\pi m \cos(m\psi)\over 2c \sin\tfrac{1}{2} \psi }\, d\psi.
\ee
Integrating by parts it is straightforward to show that
\bg
\Big|\int_{Am^{\alpha -1}}^\pi { \cos(m\psi)\over  \sin\tfrac{1}{2} \psi } \, d\psi\Big| \leq C \max\Big({1\over m^\alpha},{1\over m^{2\alpha-1}}\Big)\quad  m\gg 1.
\ee
Thus, taking $1/2 < \alpha < 1$, 
\bg
E_{\mathrm{ext}} \approx {\Gamma^2 m\over 2}\bigg[ \int_\eps^{Am^\alpha}{b^2\cos\psi+c^2\over ({4b^2\sin^2\tfrac{1}{2} \psi +c^2\psi^2})^{1/2} } \, d\psi+{1\over 2}\int_{Am^{\alpha -1}}^\pi { c\cos\psi\over  \sin\tfrac{1}{2} \psi  }\, d\psi\bigg],\quad m\gg 1.
\ee
It is preferable to add and subtract a term to obtain
\begin{align}E_{\mathrm{ext}}  \approx {\Gamma^2 m\over 2}\bigg[ \int_\eps^\infty &  \bigg(\, {b^2\cos\psi+c^2\over ({4b^2\sin^2\tfrac{1}{2} \psi +c^2\psi^2})^{1/2}}-{{b^2}\cos\psi +c^2\over c\psi}\,\bigg) \,d\psi
\notag\\
& -{b^2\over c}\, {\rm Ci}(\eps)+{c\over 2}\int_{\eps
/ m}^\pi { \cos\psi\over  \sin\tfrac{1}{2} \psi  }\, d\psi\bigg],\quad m\gg 1.
 \end{align} 
For a constant core vorticity we have an internal energy
\bg
E_{\mathrm{int}} = {m\Gamma^2\sqrt{b^2+c^2}\over 8}
\ee
to leading order.

\section*{Acknowledgements} 
ADG is very grateful to the Leverhulme Trust for a Research Fellowship which supported this research and his continuing collaboration with SC. 
We have also benefitted from discussions with K.R. Sreenivasan.

\end{document}